\documentclass[twocolumn]{aastex63}

\newcommand{\micro}{\text{$\mu$}}

\newcommand{\centi}{\text{c}}

\newcommand{\GHz}{\text{GHz}}

\newcommand{\jansky}{\text{Jy}}
\newcommand{\ujy}{\micro\text{Jy}}
\newcommand{\meter}{\text{m}}
\newcommand{\parsec}{\text{pc}}

\newcommand{\kev}{\text{keV}}

\newcommand{\ergs}{\text{erg s$^{-1}$}}

\newcommand{\E}[1]{\times10^{#1}}
\newcommand{\Msol}{{\rm M}_{\odot}}

\newcommand{\U}[3]{$#1_{-#3}^{+#2}$}

\newcommand{\beam}{\text{beam$^{-1}$}}

\newcommand{\arcs}{\ensuremath{^{\prime\prime}}}
\newcommand{\mins}{\ensuremath{.\hspace{-0.8mm}^{\prime}}}

\newcommand{\barcs}{\ensuremath{.\hspace{-1.1mm}^{\prime\prime}}}

\graphicspath{{./}{figures/}}
\usepackage[utf8]{inputenc}
\usepackage[T1]{fontenc}

\shorttitle{New Compact Binaries in Terzan 5}
\shortauthors{Urquhart et al.}

\begin{document}

\title{The MAVERIC Survey: New compact binaries revealed by deep radio continuum observations of the Galactic globular cluster Terzan 5}

\correspondingauthor{R.\ Urquhart}
\email{urquha20@msu.edu}
\author[0000-0003-1814-8620]{Ryan Urquhart}
\affiliation{Center for Data Intensive and Time Domain Astronomy, Department of Physics and Astronomy, Michigan State University, East Lansing MI 48824, USA}
\author[0000-0003-2506-6041]{Arash Bahramian}
\affiliation{International Centre for Radio Astronomy Research, Curtin University, GPO Box U1987, Perth, WA 6845, Australia}
\author[0000-0002-1468-9668]{Jay Strader}
\affiliation{Center for Data Intensive and Time Domain Astronomy, Department of Physics and Astronomy, Michigan State University, East Lansing MI 48824, USA}
\author[0000-0002-8400-3705]{Laura Chomiuk}
\affiliation{Center for Data Intensive and Time Domain Astronomy, Department of Physics and Astronomy, Michigan State University, East Lansing MI 48824, USA}
\author[0000-0001-5799-9714]{Scott M.\ Ransom}
\affiliation{National Radio Astronomy Observatory, 520 Edgemont Road, Charlottesville, VA 22903, USA}
\author{Yuankun Wang}
\affiliation{National Radio Astronomy Observatory, 520 Edgemont Road, Charlottesville, VA 22903, USA}
\author[0000-0003-3944-6109]{Craig O.\ Heinke}
\affiliation{Department of Physics, CCIS 4-183, University of Alberta, Edmonton, AB T6G 2E1, Canada}
\author[0000-0003-4553-4607]{Vlad Tudor}
\affiliation{International Centre for Radio Astronomy Research, Curtin University, GPO Box U1987, Perth, WA 6845, Australia}
\author[0000-0003-3124-2814]{James C.~A.\ Miller-Jones}
\affiliation{International Centre for Radio Astronomy Research, Curtin University, GPO Box U1987, Perth, WA 6845, Australia}
\author[0000-0003-3906-4354]{Alexandra J.\ Tetarenko}
\affiliation{East Asian Observatory, 660 N. A'oh\={o}k\={u} Place, University Park, Hilo, Hawaii 96720, USA}
\author{Thomas J. Maccarone}
\affiliation{Physics Department, Texas Tech University, PO Box 41051, Lubbock, TX 79409, USA}
\author[0000-0001-6682-916X]{Gregory R.\ Sivakoff}
\affiliation{Department of Physics, CCIS 4-183, University of Alberta, Edmonton, AB T6G 2E1, Canada}
\author[0000-0003-0286-7858]{Laura Shishkovsky}
\affiliation{Center for Data Intensive and Time Domain Astronomy, Department of Physics and Astronomy, Michigan State University, East Lansing MI 48824, USA}
\author{Samuel J. Swihart}
\affiliation{Center for Data Intensive and Time Domain Astronomy, Department of Physics and Astronomy, Michigan State University, East Lansing MI 48824, USA}
\author[0000-0002-4039-6703]{Evangelia Tremou}
\affiliation{LESIA, Observatoire de Paris, CNRS, PSL, SU/UPD, Meudon, France}




\begin{abstract}
Owing to its massive, dense core, Terzan\,5 has the richest population of millisecond pulsars known among Galactic globular clusters. Here we report new deep $2-8\,$GHz radio continuum observations of Terzan\,5 obtained with the Karl G.\ Jansky Very Large Array. We have identified a total of 24 sources within the cluster half-light radius, including 17 within the core radius. 19 are associated with previously studied millisecond pulsars and X-ray binaries. Three of the new radio sources have steep radio spectra and are located within the cluster core, as expected for millisecond pulsars. These three sources have hard X-ray photon indices ($\Gamma=1.3-1.5$) and highly variable X-ray emission, suggesting they are binary millisecond pulsars belonging to the spider class. For the most X-ray luminous of these sources, the redback spider classification is confirmed by its X-ray light curve, which shows an orbital period of 12.32 hr and double peaked structure around X-ray maximum. The likely discovery of bright binary millisecond pulsars in a well-studied cluster like Terzan\,5 highlights how deep radio continuum imaging can complement pulsar search and timing observations in finding probable eclipsing systems. The other new radio source in the core has a flat radio spectrum and is X-ray faint ($L_X \approx 2\times 10^{31}$ erg s$^{-1}$) with a photon index $\Gamma=2.1\pm0.5$, consistent with the properties expected for a quiescent stellar-mass black hole X-ray binary.
\end{abstract}

\section{Introduction}

Millisecond pulsars (MSPs) are old neutron stars that have been spun up and recycled by accreting from low-mass companions \citep{Bhattacharya91, Lorimer08}. They are the products of the evolution of low-mass X-ray binaries (LMXBs), in which mass and angular momentum are transferred to the neutron star, spinning it up to short spin periods, $\lesssim30$\,ms. Most MSPs are highly stable and hence exceptionally precise clocks, making them excellent laboratories for tests of general relativity and fundamental physics \citep[e.g.,][]{2014Natur.505..520R,2020NatAs...4...72C}. They are also unique tracers of stellar dynamical processes that efficiently generate compact binaries, possibly including neutron star--neutron star or neutron star--black hole binaries (e.g., \citealt{Clausen13,Ye20}).

Old, dense stellar environments, like globular clusters, are ideal factories for producing such exotic sources. The high interaction rates give rise to enhanced formation of binary systems, a factor of $\sim100$ greater per unit mass than in the field \citep{1975Natur.253..698K,Verbunt87}. 
The Galactic globular cluster Terzan\,5 is an exemplary target for pulsar searches. It has the highest number of confirmed MSPs of any cluster \citep{2017ApJ...845..148P}, with many more theorized to remain undetected \citep{2000ApJ...536..865F, 2011MNRAS.418..477B}. Terzan\,5 is a massive cluster ($\sim10^6\,\Msol$; \citealt{2010ApJ...717..653L,2017ApJ...845..148P,2018MNRAS.478.1520B}) that resides within the Galactic bulge/bar, at a distance of $5.9\pm0.5$\,kpc \citep{2007AJ....133.1287V}. It has core and half light radii of 9\farcs6 and 43\farcs6, respectively, and a central mass density of $(1-4)\times10^{6}\,\Msol\,\parsec^{-3}$ \citep{2010ApJ...717..653L, 2017ApJ...845..148P, 2018MNRAS.478.1520B}. Due to this high density, Terzan\,5 has the highest stellar interaction rate of any Galactic globular cluster \citep{2013ApJ...766..136B}. This in turn is responsible for the high number of MSPs, and the large population of X-ray sources \citep{2006ApJ...651.1098H}. 

Previous radio observations of Terzan\,5 have focused on using single-dish radio timing surveys that search for pulsed radio emission (e.g., \citealt{1990Natur.347..650L,2000MNRAS.316..491L, 2005Sci...307..892R,2018ApJ...863L..13A,2018ApJ...855..125C}). These campaigns have been highly successful, identifying a total of 38 MSPs residing within Terzan\,5. 

However, this method may fail to detect pulsars in the tightest binaries, which have rapidly varying acceleration \citep[e.g.,][]{2003ApJ...589..911R}. Importantly, these tight binary pulsars could be some of the most interesting systems, because some might be compact neutron star--neutron star or neutron star--black hole systems. 

Other sources that can be missed by standard single-dish surveys are pulsars with close, non-degenerate companions. In these systems, a hydrogen-rich companion star is ablated by the pulsar wind. These ``spider'' systems, named for the cannibalistic nature of the primary, can be broken down into further subclasses based on the mass of the companion; ``redbacks'' have secondaries with masses 0.1--0.5$\,\Msol$ \citep[e.g.,][]{2010ApJ...722...88A,2013IAUS..291..127R,2014ApJ...783...69G,Strader19}, while ``black widows'' have much less massive companions ($\ll0.1\,\Msol$; e.g., \citealt{1988Natur.333..237F,1990Natur.347..650L,2001ApJ...561L..93F}). 
Redbacks and black widows often show evidence of periodic radio and X-ray variability or eclipses. The radio eclipses are a result of obscuration of the radio pulsar by ionized gas that extends beyond the companion star's Roche lobe \citep{1988Natur.333..237F}. The continuum radio signal is diminished due to absorption, while scattering or dispersion makes the radio pulses challenging to detect (e.g., \citealt{1992ApJ...384L..47F, 1994ApJ...422..304T,2018MNRAS.476.1968P,2020MNRAS.tmp..561P}). Some pulsars can even be completely obscured \citep[e.g., ][]{2000ApJ...535..975C,2020MNRAS.493.6033Z}, with the radio eclipse sometimes lasting the full orbit \citep{1991ApJ...379L..69T}. The X-ray emission in these binaries is thought to originate from an intrabinary shock,
and to vary throughout the orbit due to beaming and partial eclipses as well as stochastic fluctuations \citep[see e.g., ][]{2018ApJ...861...89A,2018ApJ...869..120W,2019ApJ...879...73K}.

Currently, there are over 30 known redbacks/black widows in Galactic globular clusters\footnote{\href{http://www.naic.edu/~pfreire/GCpsr.html}{{\tt http://www.naic.edu/~pfreire/GCpsr.html}}}. A globular cluster redback (IGR J18245--2452/M28I) is one of only three confirmed transitional millisecond pulsars, systems observed to switch between a rotation-powered and accretion-powered state \citep{2013Natur.501..517P}.

\begin{figure*}[ht!]
    \centering
    \includegraphics[width=\textwidth]{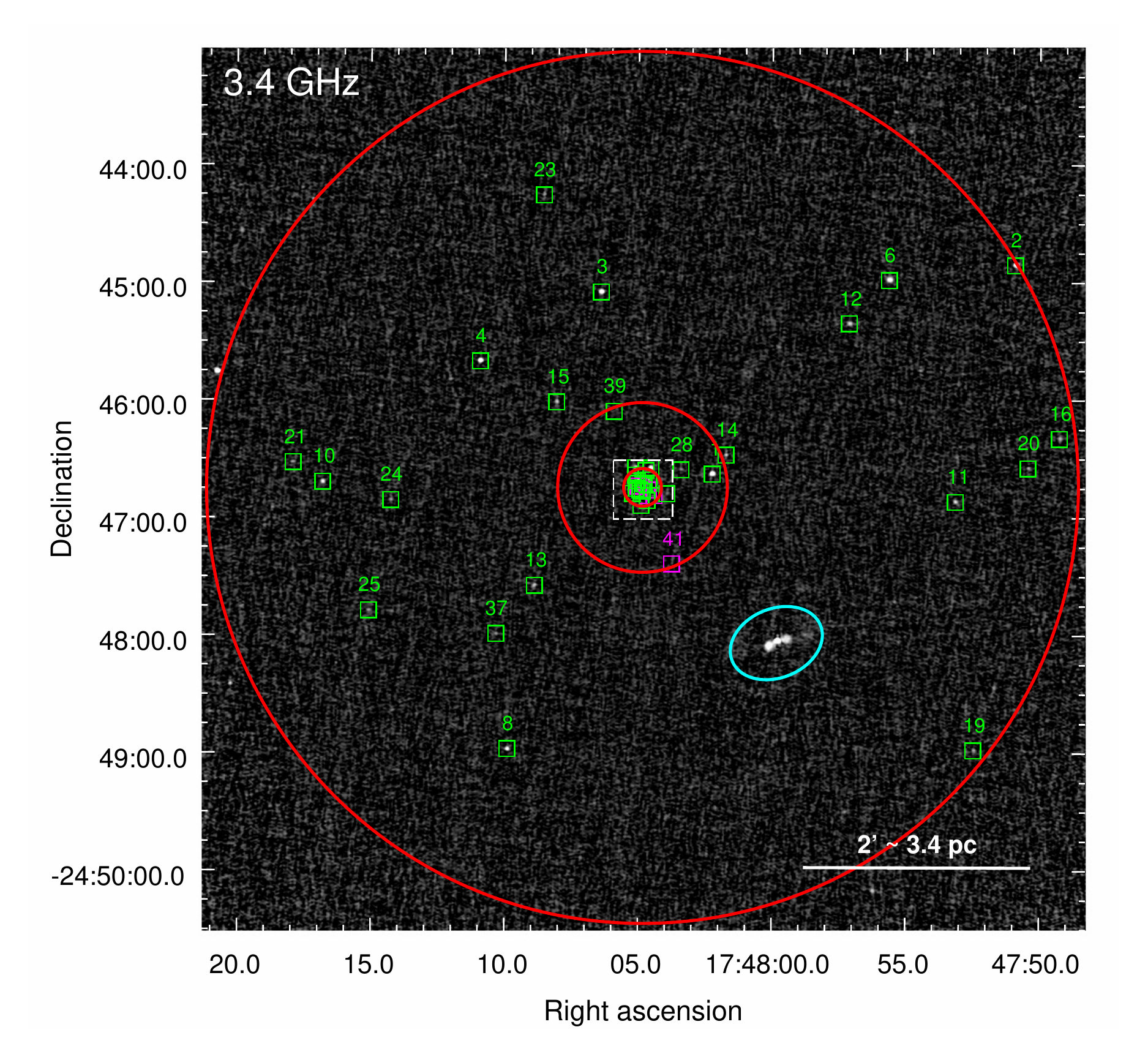}
    \caption{VLA 3.4\,GHz image centered on Terzan\,5, with a field of view of $7.5^{\prime} \times 7.5^{\prime}$. Green squares represent the positions of continuum radio sources, numbered in accordance with Table \ref{tab:radio}. Magenta squares represent the positions of new continuum radio sources within the half-light radius. Sources within the dense inner region of Terzan\,5 are not numbered for clarity, though are labeled in the subsequent Figures \ref{fig:ter5_low_freq} and \ref{fig:ter5_high_freq}. The dashed grey box indicates the area of the zoomed-in Figures \ref{fig:ter5_low_freq}, \ref{fig:ter5_high_freq}, and \ref{fig:ter5_xray}. Concentric red circles represent the core (inner) and half-light (middle) radii and the limit of our search radius (outer). The background radio galaxy excised from our sample is highlighted with the cyan ellipse.  A total of 43 radio continuum sources were identified in at least one image.}
    \label{fig:ter5_sbd_full}
\end{figure*}

Detecting the time-integrated radio continuum emission in high-resolution imaging data avoids most of these biases against detecting pulsars in close binaries. Previous continuum surveys have successfully detected sources which turned out to be new pulsars when observed with single-dish telescopes \citep[e.g., ][]{2018MNRAS.475..942F,2018ApJ...864...16M,2020MNRAS.493.6033Z}. Similarly, the first candidate globular cluster pulsar was detected via a radio continuum survey with the Very Large Array \citep{1985AJ.....90..606H} and later confirmed with an observation at the Lovell radio telescope \citep{1987Natur.328..399L}.


Existing radio continuum observations of Terzan\,5  were able to identify several bright, isolated MSPs \citep{1990ApJ...365L..63F,2000ApJ...536..865F}. However, these observations were limited by their resolution; \citet{2000ApJ...536..865F} reported a $\sim2\,$mJy diffuse steep-spectrum component 
located within the core, which was attributed to a population of unresolved  pulsars. Likewise, imaging surveys with both the NVSS (at 21\,cm; \citealt{1998AJ....115.1693C}) and Effelsberg (11 and 21\,cm; \citealt{2011A&A...532A..47C}) find evidence of extended emission. Clearly Terzan\,5 is a prime target for deep, high-resolution continuum observations,
to determine whether the MSPs discovered from timing observations can account for 
all of this radio flux, or whether other sources are present.

Radio imaging data is also sensitive to compact binaries other than pulsars, including low-mass X-ray binaries that can produce compact jets. Three neutron star 
low-mass X-ray binaries in Terzan 5 are known (see discussion in \citealt{2014ApJ...780..127B}). Accreting stellar-mass black holes, if present, would also produce observable compact radio jets. It was long argued that most, if not all, stellar-mass black holes had been ejected from present-day globular clusters due to gravitational interactions \citep{1993Natur.364..421K, 1993Natur.364..423S}. But recently, evidence has been mounting---both observationally \citep{2007Natur.445..183M,2012Natur.490...71S,2013ApJ...777...69C,2015MNRAS.453.3918M,2018MNRAS.475L..15G, 2018ApJ...855...55S, 2019MNRAS.485.1694D,2019A&A...632A...3G} and theoretically \citep{2013MNRAS.430L..30S,2015ApJ...800....9M,2016MNRAS.462.2333P,2018ApJ...864...13W}---that a significant population of black holes formed and remains within clusters. The abundance of black holes in present-day globular clusters places important constraints on the formation of gravitational wave sources and on the long-term dynamical evolution of clusters themselves (e.g., \citealt{2016PhRvD..93h4029R,2019ApJ...871...38K}).

Here we present a deep radio survey of Terzan\,5 with the Karl G.\ Jansky Very Large Array (VLA) at frequencies 2--8 GHz. In Section \ref{sec:data_analysis} we outline our data analysis process. In Section \ref{sec:results} we detail our findings, comparing our population of radio sources to previous work, and discuss new candidate MSPs and a candidate black hole X-ray binary. We summarize our results in Section \ref{sec:conclusion}.

\section{Data analysis} \label{sec:data_analysis}
\subsection{Radio observations}

\begin{figure*}
    \centering
    \includegraphics[width=0.48\textwidth]{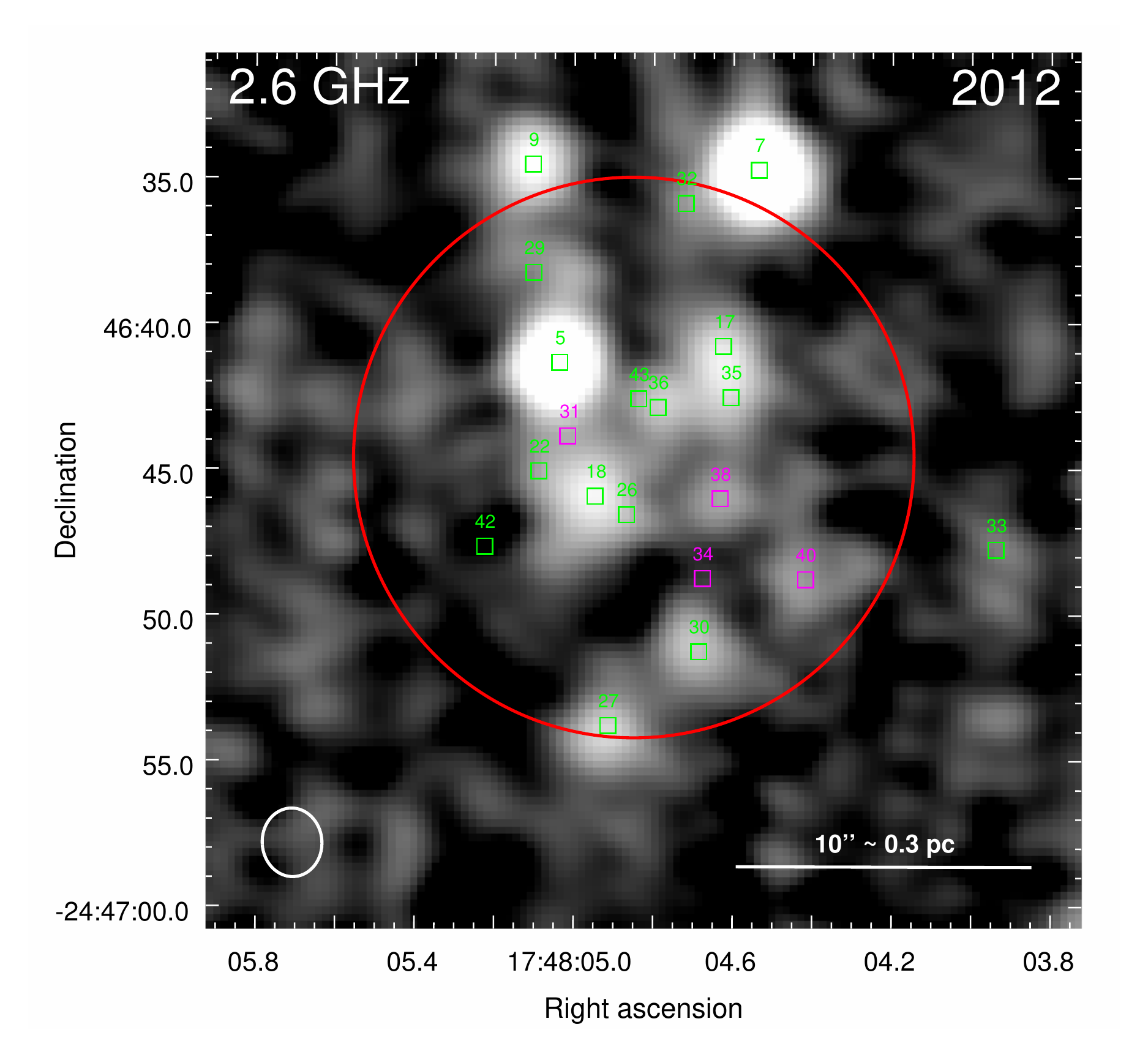}
    \includegraphics[width=0.48\textwidth]{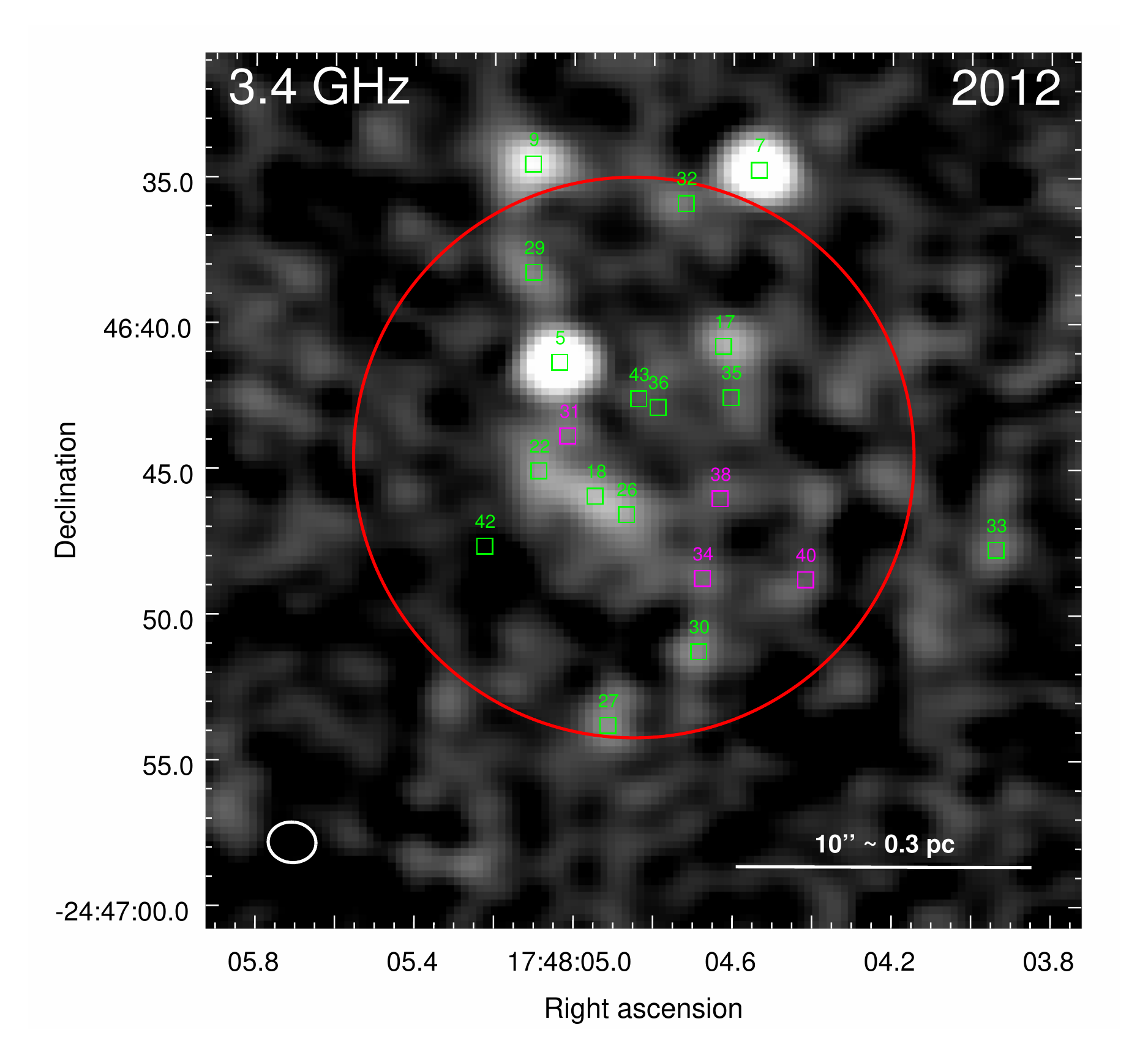}\\
    \includegraphics[width=0.48\textwidth]{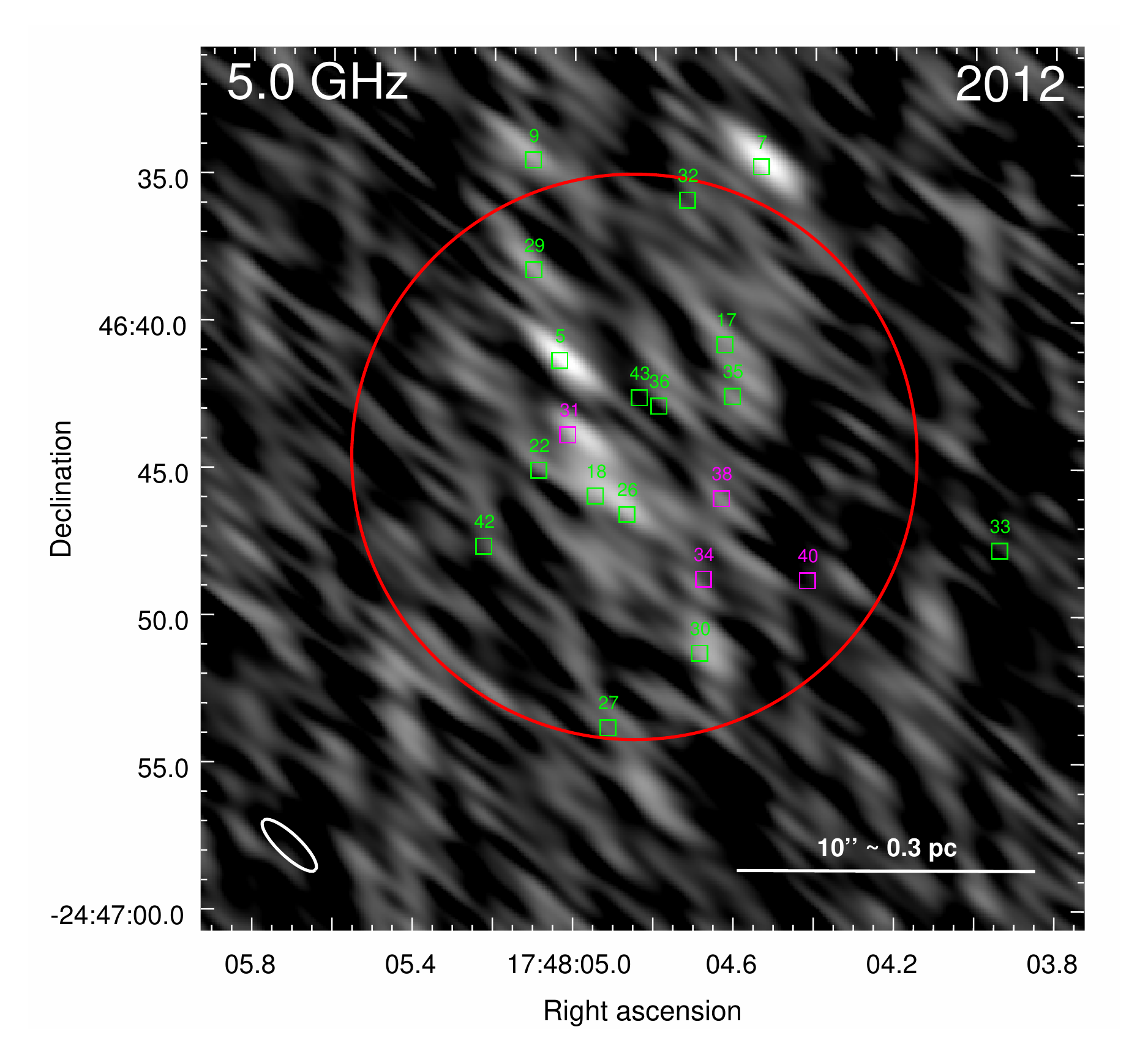}
    \includegraphics[width=0.48\textwidth]{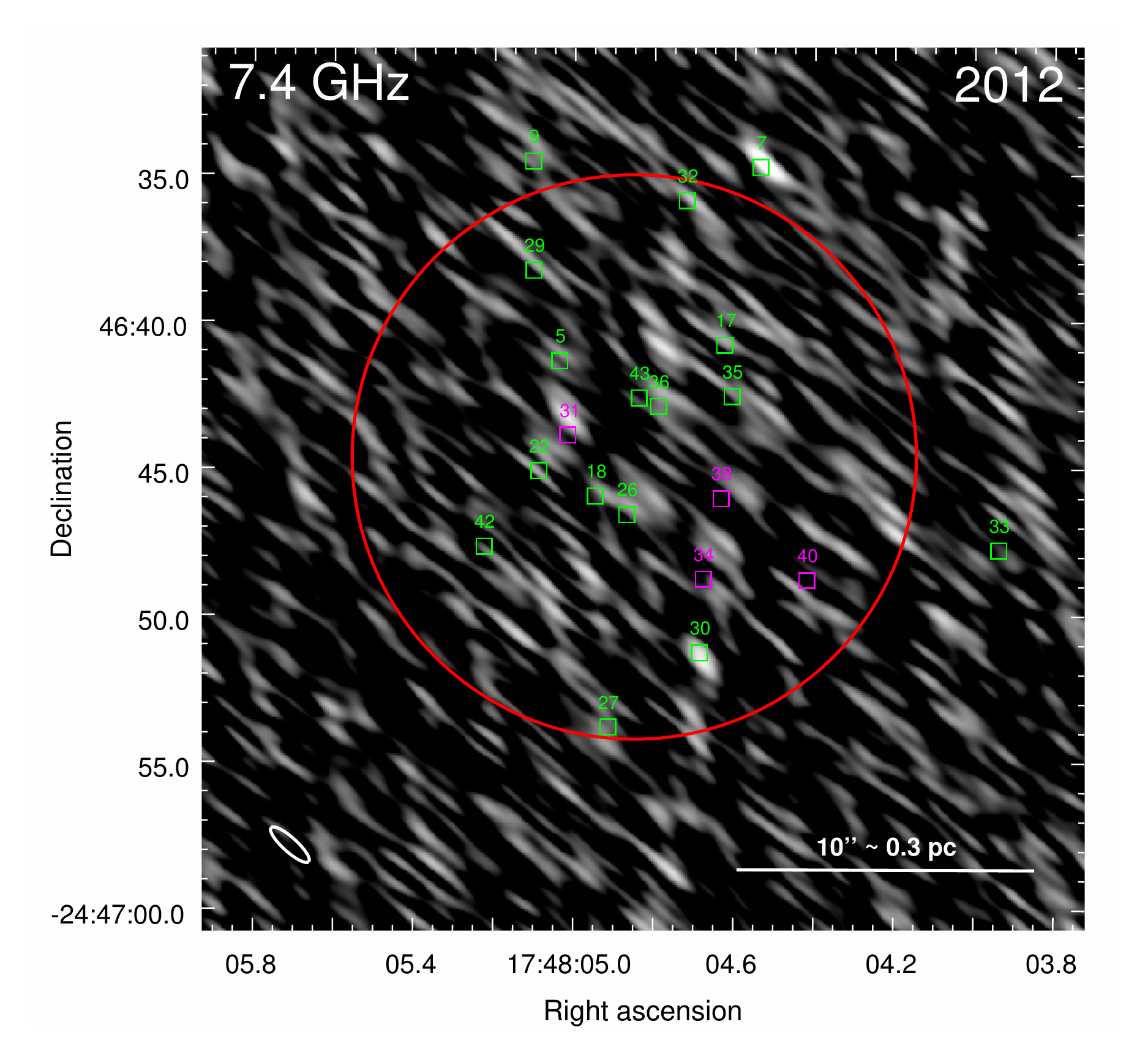}
    \caption{$0.5^{\prime} \times 0.5^{\prime}$ zoom-in of Figure \ref{fig:ter5_sbd_full}, focusing on the core of Terzan\,5. We show the quasi-simultaneous 2012 VLA observations: 2.6\,GHz top left; 3.4\,GHz top right; 5.0\,GHz bottom left; 7.0\,GHz bottom right. The red circle in each panel denotes the core radius of Terzan\,5. Green and magenta boxes represent known and new continuum sources, respectively. White ellipses at the bottom left indicate the synthesized beam sizes of each image.}
    \label{fig:ter5_low_freq}
\end{figure*}

\begin{figure*}
    \centering
    \includegraphics[width=0.48\textwidth]{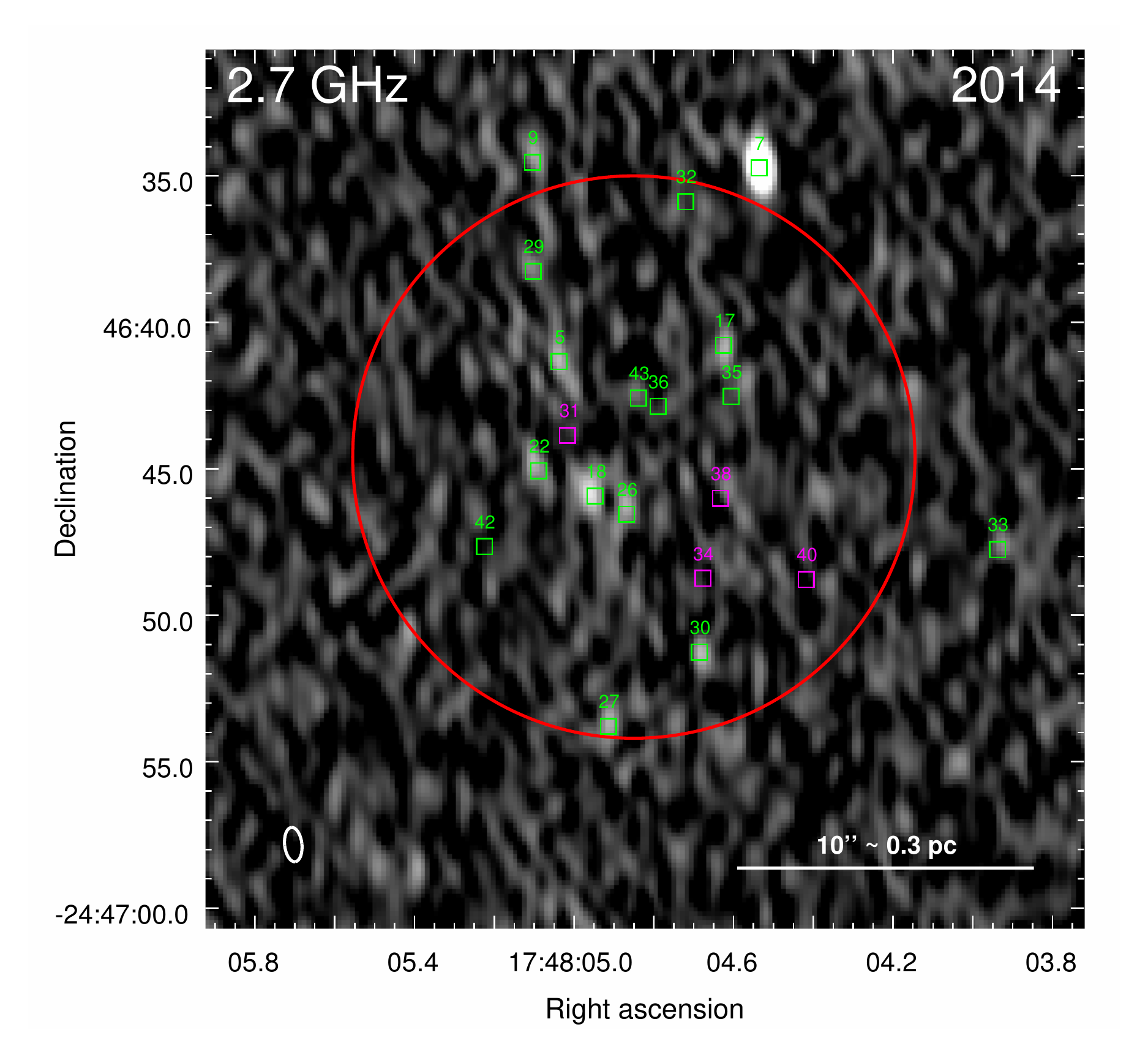}
    \includegraphics[width=0.48\textwidth]{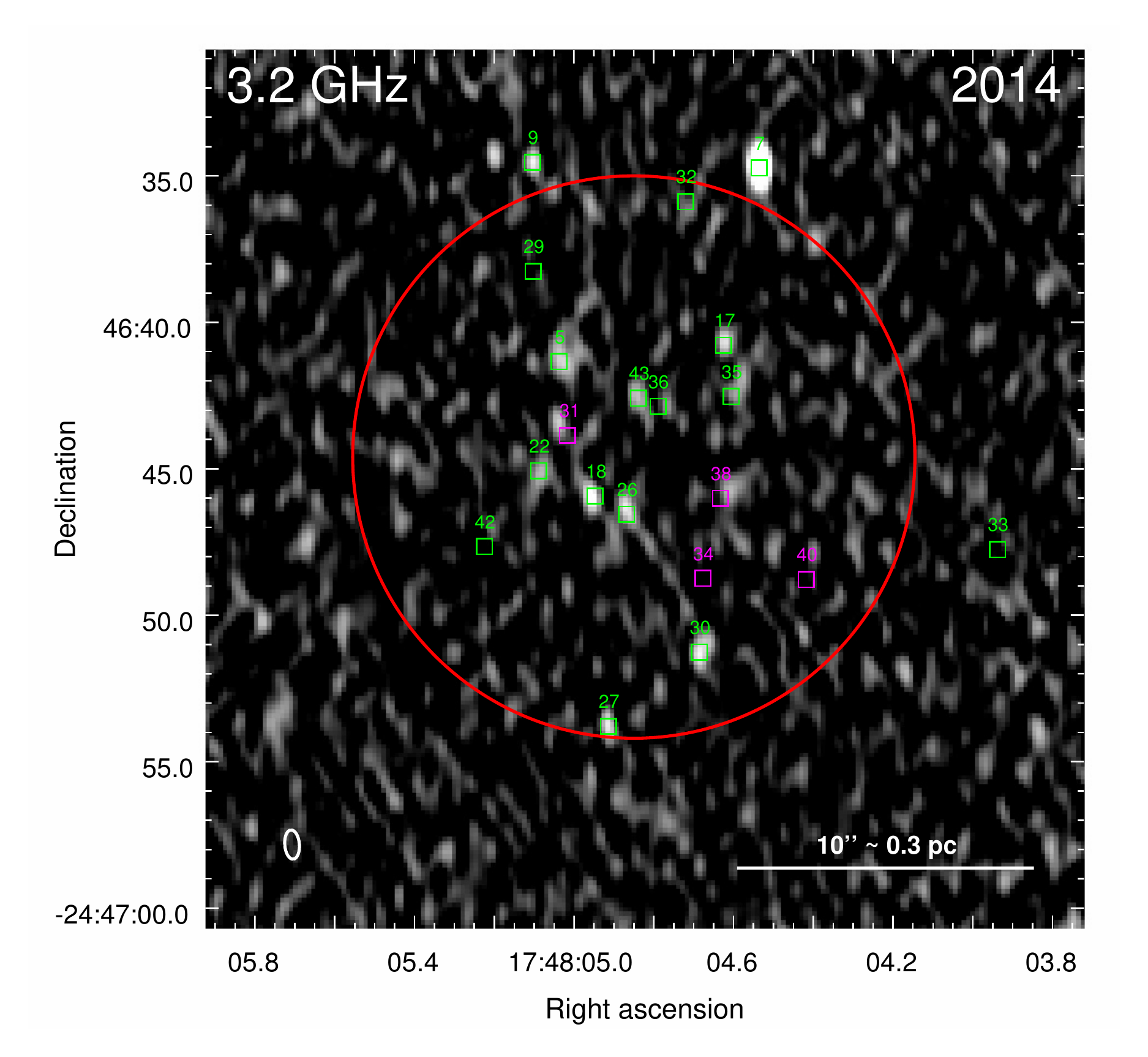}\\
    \includegraphics[width=0.48\textwidth]{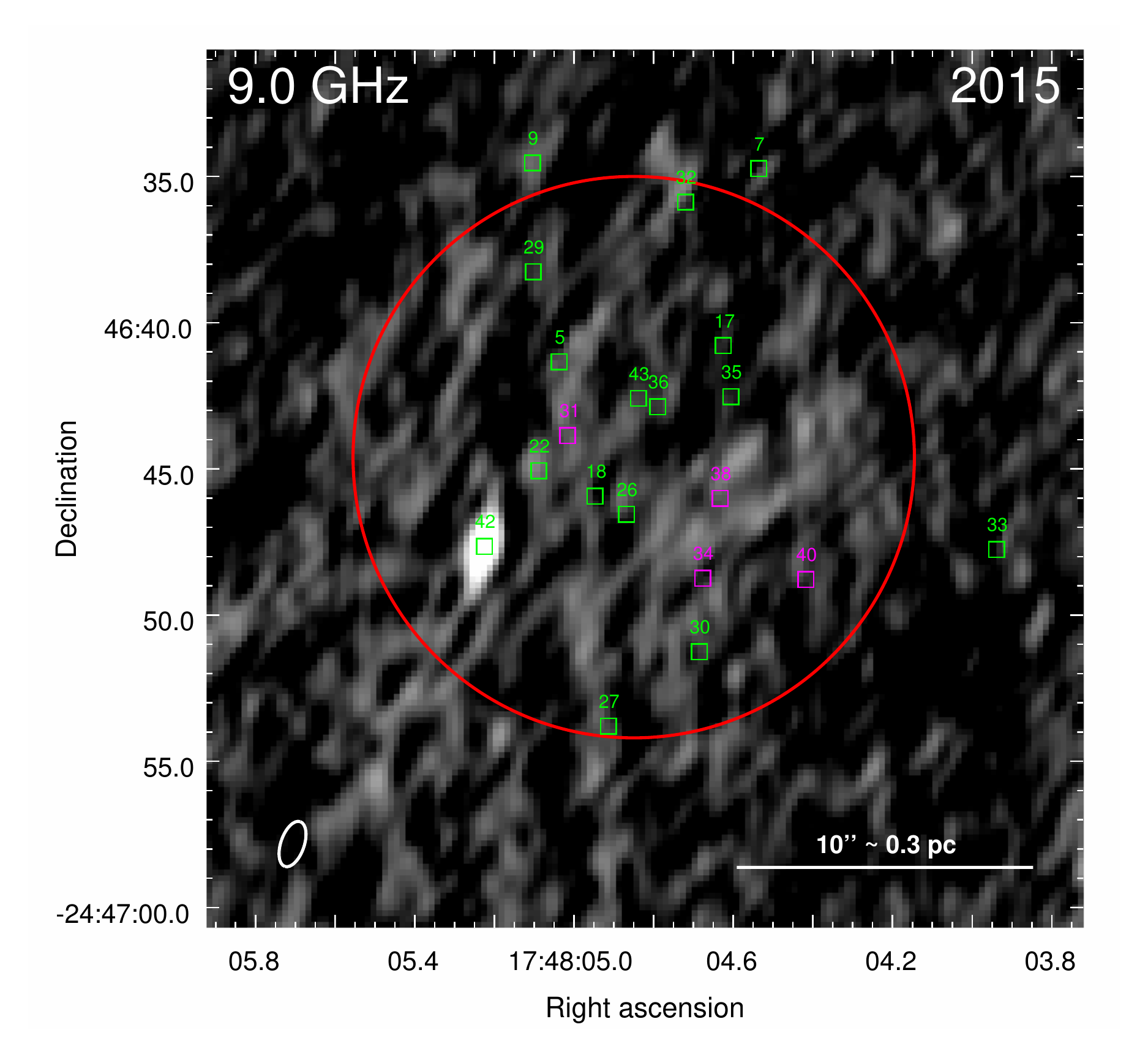}
    \includegraphics[width=0.48\textwidth]{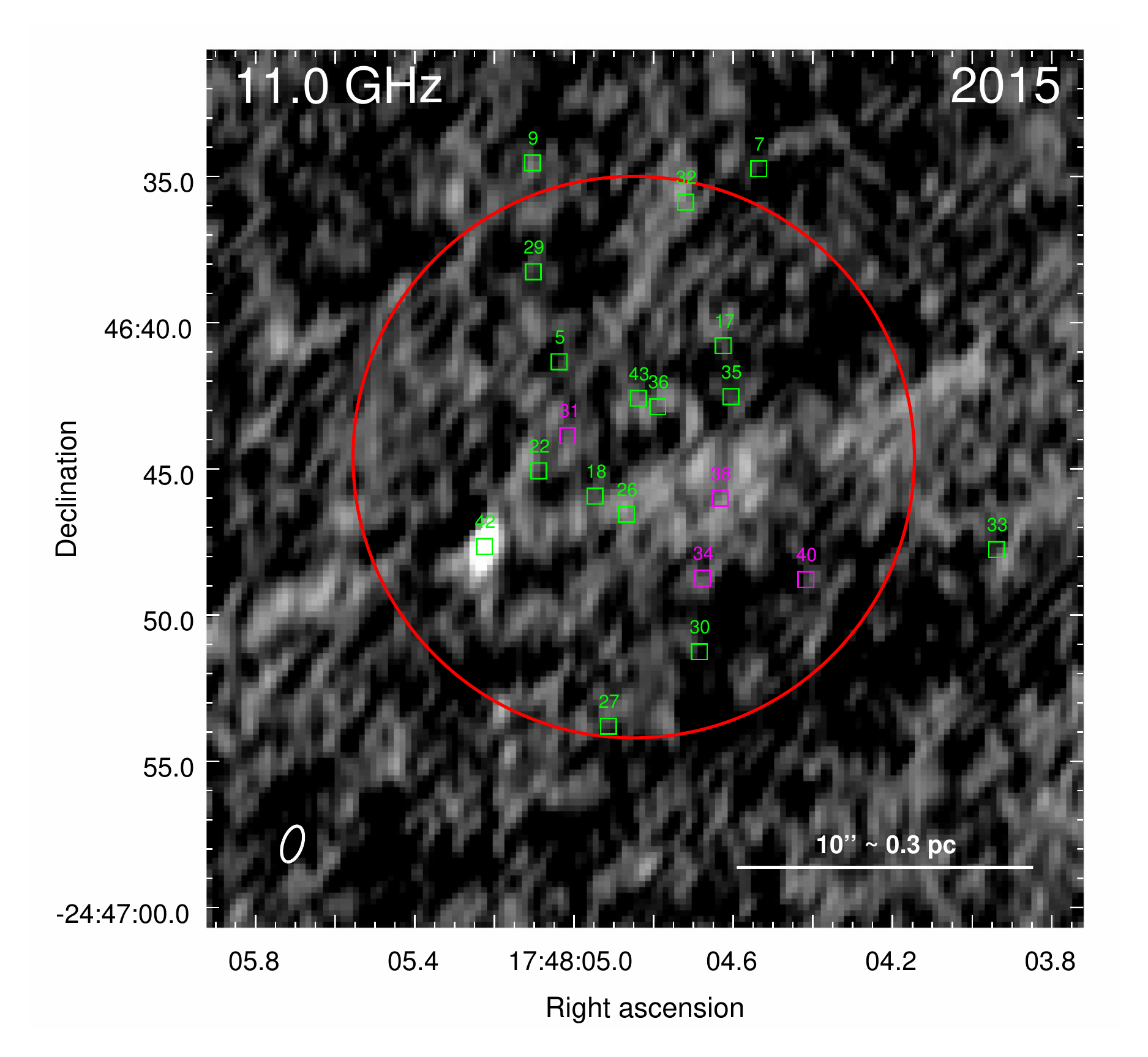}
    \caption{As in Figure \ref{fig:ter5_low_freq}, but for the  2014 and 2015 observations: 2014 2.7\,GHz top left; 2014 3.2\,GHz top right; 2015 9.0\,GHz bottom left; 2015 11.0\,GHz bottom right. The 2014 data are used only for position information.}
    \label{fig:ter5_high_freq}
\end{figure*}

In 2012, we observed Terzan\,5 with the VLA as part of program VLA/12B-073 (PI J.\ Strader) and the MAVERIC survey, a project designed to study radio sources in globular clusters (Shishkovsky et al.\ 2020, in prep.; Tudor et al.\ 2020, in prep.). Unlike all other clusters in that survey, which were only observed at C band (4--8 GHz) for maximal sensitivity to flat spectrum sources, we split the observations of Terzan 5 between C band and S band (2--4 GHz) because of the known large population of millisecond pulsars.

C-band observations were taken on 2012 September 10--12, for a total observing time of 5 hours. The data were split into two 1.024\,GHz basebands with 8-bit sampling, centered at 5.0 and 7.4\,GHz. S band observations were taken on 2012 September 13--14, again for a total of 5 hours and similarly split into two 1.024\,GHz basebands,  centered at 2.5 and 3.5\,GHz (after flagging, the final central frequencies were 2.6 and 3.4 GHz). For both C and S band observations the VLA was in the hybrid BnA configuration, which produces B configuration-like resolution with a more nearly circular beam for southern sources. J1751$-$2524 was used as the phase calibrator, while 3C48 and 3C286 were used as the flux/bandpass calibrators for C band and S band, respectively.

We also used additional archival observations of Terzan\,5 taken as part of the VLA pulsar test program TPUL001, albeit selectively. A total of 3.5\,hr of S band observations were conducted on 2014 February 14 while the VLA was moving between the BnA and A configurations. These data were split into two basebands (each 512\,MHz wide) and centered at 2.7 and 3.2\,GHz. 3C286 was used as the flux/bandpass calibrator, while J1751$-$2524 was used as the phase calibrator. We found that the non-standard, experimental setup of these observations led to discrepancies in the flux calibration compared with other datasets. However, they still have the advantage of being higher resolution than the 2012 BnA S band data. Hence these 2014 data were only used to help distinguish closely spaced or faint sources from each other in the crowded core of the cluster; no flux densities from these data were used.

\begin{deluxetable}{ccccc}
\tabletypesize{\footnotesize}
\tablecaption{Summary of VLA observations of Terzan 5.}
\tablehead{
\colhead{Freq} & \colhead{rms} & \colhead{Beam FWHM} & \colhead{Beam PA} & \colhead{Obs Date}\\
\colhead{(GHz)} & \colhead{($\mu$Jy bm$^{-1}$)} & \colhead{($^{\prime\prime} \times ^{\prime\prime}$)} & \colhead{(deg)} &
}
\startdata
2.6 & 6.3 & $2.35\times2.05$ & 4.0 & 2012 Sep 13--14 \\
3.4 & 4.0 & $1.64\times1.39$ & 84.4 & 2012 Sep 13--14 \\
5.0 & 4.3 & $2.44\times0.75$ & 46.8 & 2012 Sep 10--12 \\
7.4 & 3.5 & $1.73\times0.48$ & 48.1 & 2012 Sep 10--12 \\
\hline
2.7 & 5.2 & $1.18\times0.59$ & 6.2 & 2014 Feb 14 \\
3.2 & 5.8 & $1.00\times0.50$ & 4.6 & 2014 Feb 14 \\
9.0 & 5.0 & $1.62\times0.80$ & -19.0 & 2015 Mar--Apr \\
11.0 & 5.8 & $1.28\times0.66$ & -19.2 & 2015 Mar--Apr\\
\enddata
\end{deluxetable}
\label{tab:obs}

Finally, we include additional shallower high-frequency archival VLA observations of Terzan\,5. Between 2015 March and April, a total of 2.25 hours of  X band observations of Terzan\,5 were taken in order to study the transient low-mass X-ray binary EXO 1745$-$248 (Project code: 14B-216; \citealt{2016MNRAS.460..345T}). The VLA was in B configuration. These observations comprised of two 2.048\,GHz basebands with 3-bit sampling centered on 9 and 11\,GHz. 3C286 was used as both the bandpass and flux calibrator. J1751$-$2524 was used as the phase calibrator. For full details on the observations, see \citet{2016MNRAS.460..345T}. All the radio 
observations are listed in Table \ref{tab:obs}.

Flagging, calibration, and imaging were performed using standard tasks in the Astronomical Image Processing System (AIPS; \citealt{2003ASSL..285..109G}). For each observation, the basebands were imaged separately, resulting in a total of 8 images taken from 2012--2015. Each image was corrected for the primary beam sensitivity at its central frequency. Table \ref{tab:obs} summarizes the properties of the final images. The final 3.4 GHz image is shown in Figure \ref{fig:ter5_sbd_full}, and core zoom-ins in Figures \ref{fig:ter5_low_freq} and \ref{fig:ter5_high_freq}.

Astrometric corrections were applied to the 2012 observations. Using several, point-like background sources at the edges of the images, we find that a slight ($\sim1^{\circ}$) rotation is required to align the 2012 images to the 2014 and 2015 images. Sources within the core are close to the rotation point and thus not meaningfully affected, while sources at the half-light radius (43\farcs8) will be shifted by $\sim0$\farcs1.

\subsection{Source finding}
Source finding is performed using the Python Blob Detection and Source Finder ({\sc PyBDSF}; \citealt{2015ascl.soft02007M}) software package, version 1.9.0rc. We search all primary-beam corrected images independently for radio sources. The source extraction software estimates the mean background and rms maps using a sliding box, set to the PyBDSF default pixel dimensions and step increment. In each image, we search within the cluster half-light radius ($r<0\mins73$) and select sources with a signal-to-noise (S/N) ratio of $>3$. Beyond the half-light radius but within the half-power point at 6 GHz ($r=0\mins73-3\mins7$) we select sources with a S/N $> 5$, using a higher threshold to minimize the chance of false detections over this larger region. We do not analyze sources beyond the 50\% primary beam at 6 GHz (r>$3\mins7$), equivalent to $\sim4$ half-light radii from the cluster core.

Because we are interested in unresolved stellar sources, we measure flux densities assuming all objects are point sources. In {\sc PyBDSF}, we fit detected sources with Gaussians that are fixed to the dimensions of the image synthesized beam. One source that is clearly a double-lobed AGN (RA = 17:47:59.817, Dec = $-24$:48:02.46, cyan ellipse in Figure \ref{fig:ter5_sbd_full}) is manually removed. 

As suggested by the VLA Observational Status Summary\footnote{https://science.nrao.edu/facilities/vla/docs/manuals/oss/per formance/positional-accuracy}), we use 10\% of the FWHM of the synthesized beam as a conservative estimate of the positional uncertainties. Given the fact that previous pulsar timing surveys have determined precise positions for many of these radio sources (see Section \ref{sec:match}), we have the fortunate chance to independently test our astrometry. We find no systematic offset between the continuum and timed positions of the sample of high signal-to-noise and non-blended continuum sources associated with known pulsars. Reassuringly, there is a rms scatter of $\approx0$\farcs2 in each of R.A. and Dec, consistent with the $1\sigma$ positional uncertainties estimated from the synthesized beam.

We cross-match sources across all images. Sources are considered to match if the respective 3$\sigma$ positional error ellipses overlap. We find a total of 43 radio sources, with 37 of them being detected in at least two different images. The final ICRS R.A. and Dec. of each source are calculated using the variance-weighted mean position and listed in Table \ref{tab:radio}. The final positional uncertainties for each source come from the highest resolution image (if the source is detected in multiple images) in which the source is detected with confidence. The source ID numbers given in Table \ref{tab:radio} are used to refer to our radio continuum sources hereafter as Ter5-VLA{\tt XX} (where {\tt XX} is the ID number). For each significant source listed in the catalog, we also report corresponding upper limits for any images in which it was not detected. These were calculated as 3 times the local noise as measured from the rms images.

\subsection{Radio spectral analysis}

We calculate the spectral index $\alpha$ (defined as $S_{\nu} \propto \nu^{\alpha}$) of each radio source using the 2012 quasi-simultaneous 2.6, 3.4, 5.0 and 7.4\,GHz flux densities. For non-detections, 3$\sigma$ upper limits are used. The 9.0 and 11.0\,GHz flux densities are ignored in the spectral analysis due to their non-simultaneity. 

For each source, the spectral index is modeled using the Bayesian Markov Chain Monte Carlo software \verb|JAGS| \citep{2012ascl.soft09002P} assuming a power law model and a uniform prior on $\alpha$ between $-$3.5 and 3.5. The modeling self-consistently takes into account both measurements and upper limits. The median of the posterior distribution of the spectral indices and 1$\sigma$ uncertainties are reported in Table \ref{tab:radio}. For the few sources with only a single low-frequency measurement, a $3\sigma$ upper limit on $\alpha$ is reported instead.

\subsection{X-ray observations}

\begin{figure}[t]
    \centering
    \includegraphics[width=0.48\textwidth]{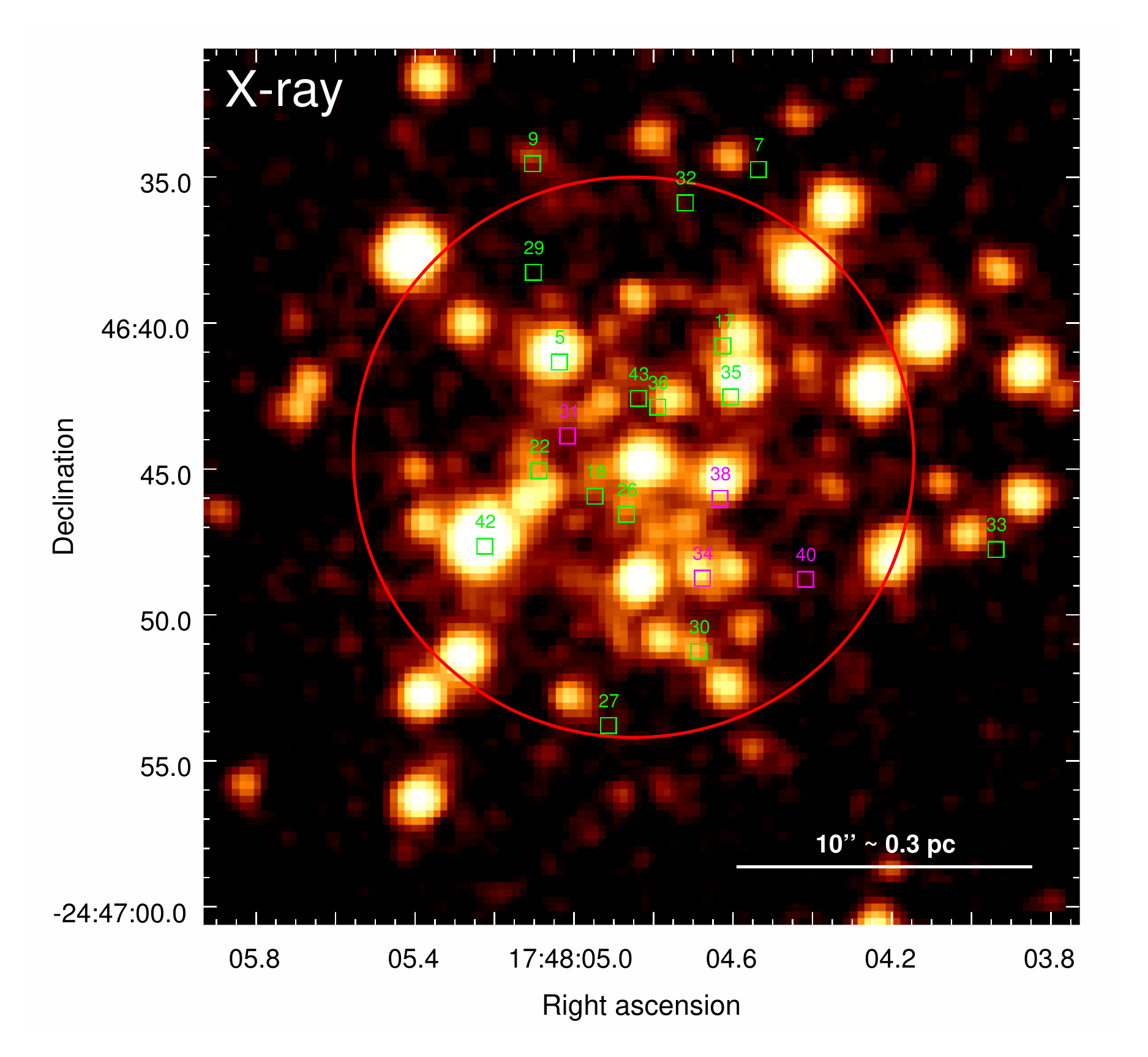}
    \caption{Stacked \textit{Chandra}/ACIS $0.5^{\prime} \times 0.5^{\prime}$ image of the core of Terzan\,5, adaptively-smoothed with a Gaussian kernel of 2 pixels for visualization purposes. Green boxes indicate the positions of continuum radio sources coincident with known pulsars; magenta boxes represent new candidate sources (Table \ref{tab:radio}). The red circle represents the core radius. A total of 22 radio continuum sources have matching X-ray counterparts.}
    \label{fig:ter5_xray}
\end{figure}

To compare the radio to X-ray properties of these sources, we use the catalog of \citet{goose}, which contains a comprehensive analysis of {\it Chandra} X-ray observations of Terzan\,5 (see also \citealt{2018ApJ...864...28B}). A total of 18 \textit{Chandra} observations were combined for a maximum exposure time of $\sim750$\,ks (some sources did not fall on the chip in every observation). The X-ray catalog contains nearly 200 sources and extends out to a projected radius of $\sim1^{\prime}$, beyond the half-light radius. The stacked \emph{Chandra}/ACIS image of Terzan\,5 is shown in Figure \ref{fig:ter5_xray}.

All X-ray sources in the catalog have been fit with an absorbed power-law model. Two absorption components are used; the first is fixed to the cluster absorption ($n_{\mathrm H} = 2.07\times10^{22}\,\centi\meter^{2}$), while the second is left free to capture the intrinsic source absorption. From this catalog, we use the following measurements: average 0.5--10\,keV unabsorbed luminosity, best-fitting power-law photon index, and a statistic measuring the variability. 22 of our radio sources have X-ray counterparts, with the X-ray properties detailed in Table \ref{tab:xray}. Sources without listed counterparts can be taken to have $L_X \lesssim 1.2 \times 10^{30}$ erg s$^{-1}$ for the assumed power-law model.

To ensure accurate fits for the new candidate radio sources with X-ray counterparts (discussed in detail in Sections \ref{sec:pul_can} and \ref{sec:bh_can}), we take this analysis a step further in carefully assessing the quality of the model fits and whether additional intrinsic source absorption is necessary. In all four cases (for (Ter5-VLA31, Ter5-VLA34, Ter5-VLA38 and Ter5-VLA40) we find that an absorbed power law provides a reasonable fit to the X-ray data with the $n_{\mathrm H}$ fixed to the foreground value. The X-ray luminosities and photon indices from these new fits are marked in Table \ref{tab:xray}.

\section{Results and discussion} \label{sec:results}

\subsection{Cross-matching with known sources} \label{sec:match}
Using deep continuum VLA observations of Terzan\,5, we have identified a total of 43 radio sources (Table \ref{tab:radio}): 17 within the core, and a further within the half-light radius but outside of the core. We matched our catalog with positions of the 38 known pulsars in Terzan\,5 (\citealt{2005Sci...307..892R,2006Sci...311.1901H, 2017ApJ...845..148P,2018ApJ...855..125C}, S. Ransom, in prep.). The uncertainties on the pulsar positions were negligible compared to the positional uncertainties on our continuum sources, so if one of our radio sources was consistent with a pulsar position within the continuum 3$\sigma$ positional uncertainty, we considered it a match. 

Of our radio sources, 17 have positions that match, within 3$\sigma$, known pulsars detected via single-dish timing studies of Terzan\,5. While crowding within the core of Terzan\,5 is a potential issue, only one continuum source (Ter5-VLA18) matches multiple pulsars (Ter5Z and Ter5ae).

In addition, our sample also includes the outbursting neutron star LMXB EXO 1745$-$248 \citep{2016MNRAS.460..345T} and the candidate transitional MSP Terzan\,5--CX1 \citep{2018ApJ...864...28B}, both of which reside in the core. This leaves a total of four new sources discovered within the core, one additional new source within the half-light radius, and 19 further new sources identified beyond the half-light radius. Only a single pulsar or LMXB in Terzan\,5 is outside the half-light radius (the pulsar Ter5J), so we expect nearly all the 19 more distant sources will be unassociated with the cluster.

\begin{figure}[t]
    \centering
    \includegraphics[width=0.45\textwidth]{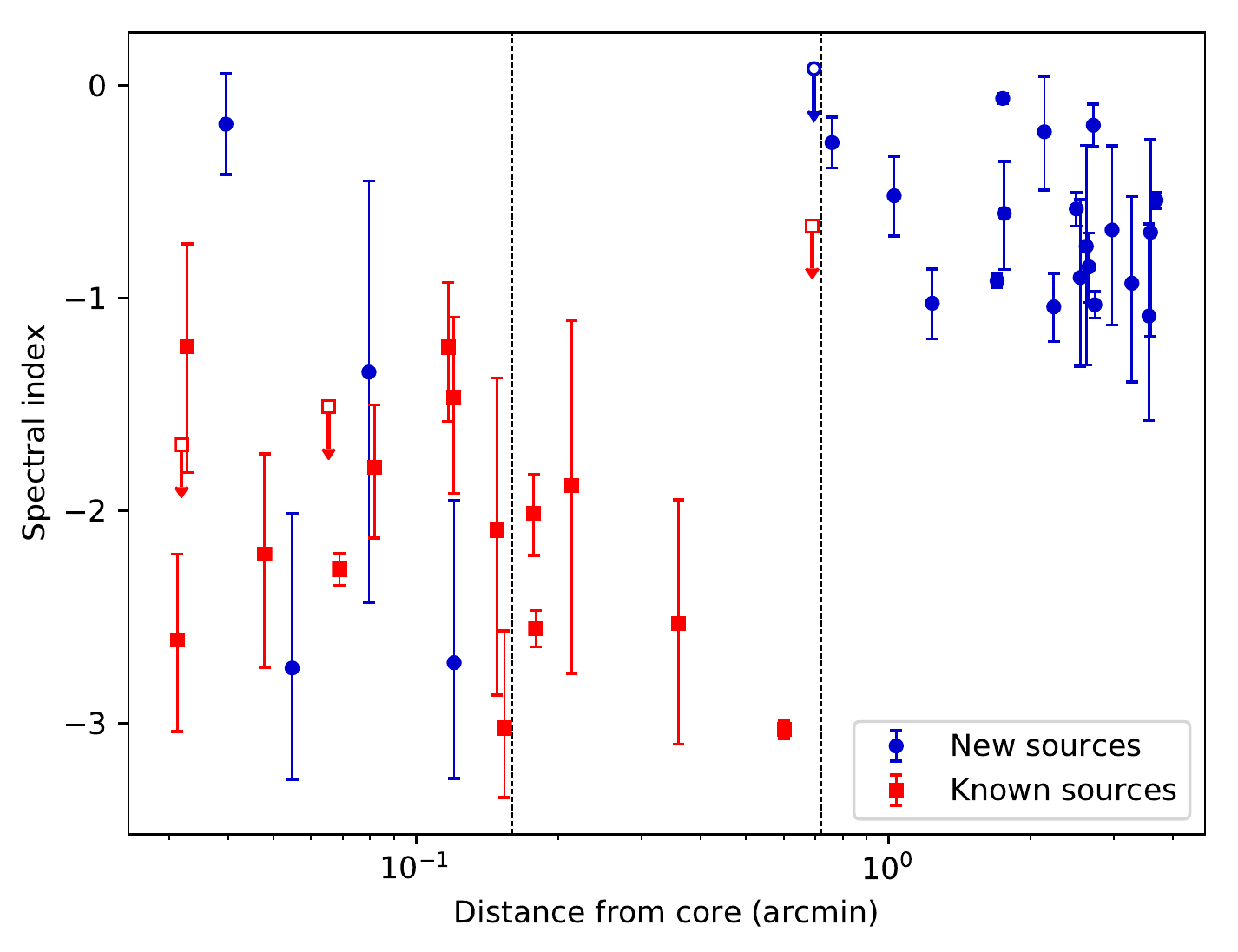}
    \caption{Radio spectral indices as a function of angular distance from the center of Terzan\,5. Red squares represent sources coincident with known objects in Terzan\,5, and blue circles represent radio continuum sources which do not have known associations. Open markers indicate 3$\sigma$ upper limits on the spectral index. Vertical dashed lines indicate the core (left) and half-light (right) radii. 41 radio sources are plotted---those with at least one flux density measured in the quasi-simultaneous images from 2012 (Table \ref{tab:radio}). The cluster of flatter-spectrum ($-1\lesssim\alpha\lesssim0$) sources outside of the half-light radius are likely background sources.}
    \label{fig:radvspec}
\end{figure}
To investigate this in more detail, we analyze the spatial distribution and radio spectral properties of our sources, comparing new and known sources (Figure \ref{fig:radvspec}). The majority of sources within the half-light radius are steep spectrum ($\alpha\lesssim-1$) and associated with known MSPs. In contrast, all distant sources (beyond the half-light radius) have, on average, much flatter spectra ($-1\lesssim\alpha\lesssim0$). These sources are likely background active galactic nuclei
and we do not investigate them further. Based on this local surface density of background sources, we expect no (0.04$\pm$0.01) background sources within the core of Terzan\,5 and around one (0.8$\pm$0.2) background source within the half-light radius. This adds to our confidence that all of the core sources are indeed Terzan\,5 members, while there are likely to be $\lesssim2$ background sources within the half-light radius. In fact, we identify a single source within the half-light radius that we consider likely to be a background object (Ter5-VLA41, discussed below).

\subsection{Comparing continuum searches with timed pulsar searches}

\begin{figure}[t]
    \centering
    \includegraphics[width=0.46\textwidth]{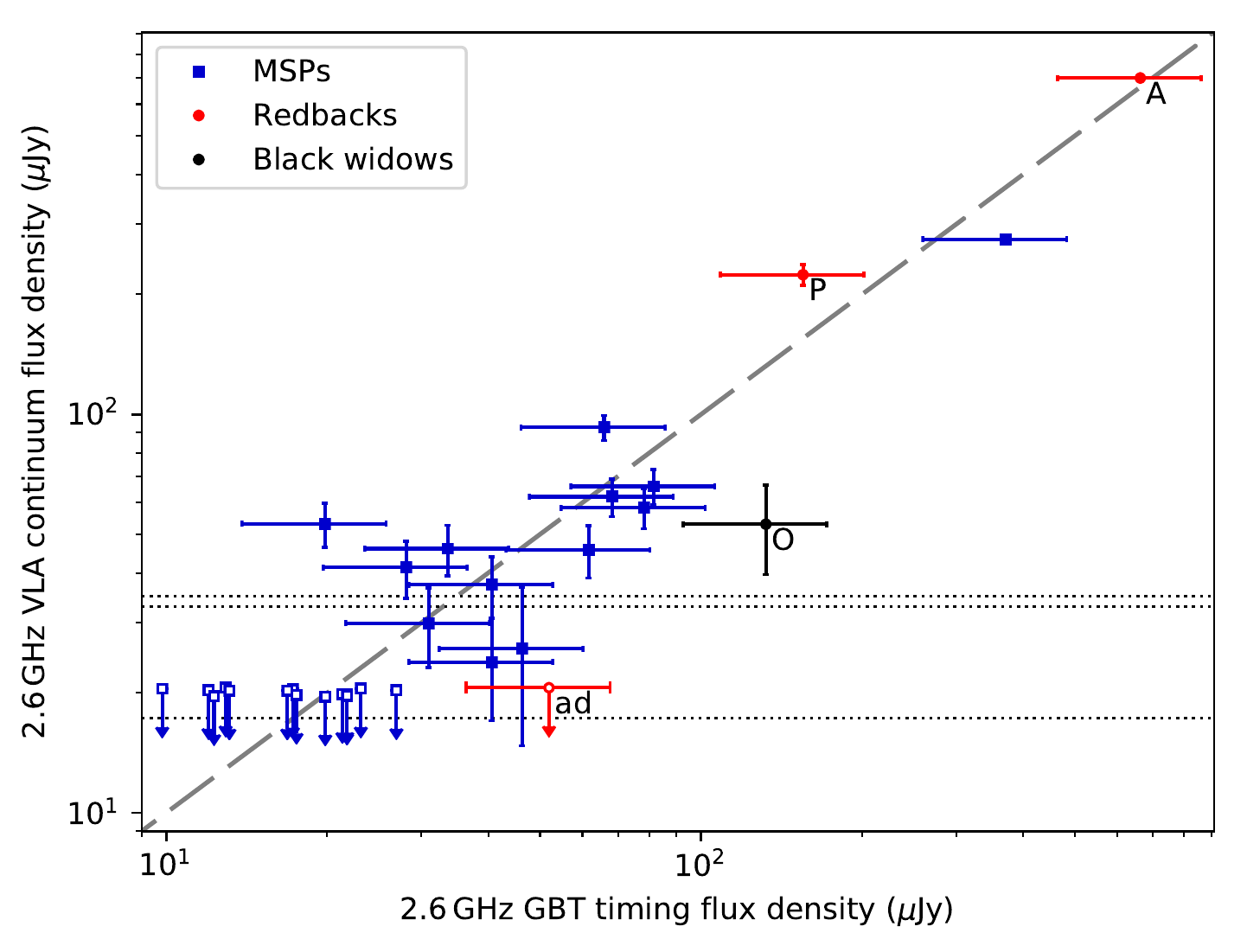}
    \caption{VLA 2.6\,GHz time-integrated flux densities of pulsars in Terzan\,5  plotted against their associated pulsar flux densities from single-dish timing analysis \citep{2005Sci...307..892R,2006Sci...311.1901H, 2017ApJ...845..148P, 2018ApJ...855..125C}. Blue points represent standard MSPs discovered in single-dish timing searches. Red and black circles represent the subset of eclipsing sources: redbacks (Ter5A, Ter5P and Ter5ad) and black widows (Ter5O), respectively. Open points are the MSPs not detected in our continuum images (3$\sigma$ upper limits are plotted). The dashed grey line is a 1:1 relation. Horizontal dotted lines indicate the flux densities of our new candidate redback pulsars (from brightest to faintest: \#38, 40, 34 in Table \ref{tab:radio}). Generally, there is agreement between the two kinds of measurements measurements. The figure also shows that our candidate redbacks are brighter than a significant fraction of known MSPs, highlighting the challenge of detecting redbacks in timing surveys.}
    \label{fig:pul_flux_comp}
\end{figure}

Currently, there are a total of 38 known MSPs in Terzan\,5 that have been discovered in single-dish timing analysis \citep{1990Natur.347..650L,2000MNRAS.316..491L, 2005Sci...307..892R,2018ApJ...863L..13A,2018ApJ...855..125C}. We detected 18 of these sources in our continuum data (two pulsars, Ter5Z and Ter5ae, are both coincident with Ter5-VLA18), leaving 20 known MSPs not identified by our survey. In this section we compare the pulsar sample detected with timing techniques to the sample detected via time-integrated continuum emission.

In Figure \ref{fig:pul_flux_comp} we compare the flux densities measured with the two different techniques. The MSP timing surveys, typically operating at 1.4 and 2.0\,GHz, report the pulsar mean flux density derived from the radiometer equation, scaled appropriately using a reference flux calibrator. We transform these flux densities from 2.0 to 2.6$\,\GHz$ using the spectral indices derived from the flux calibrated pulse profiles from the summed timing observations of each source. We assume uncertainties on these flux densities of 30\%. For Ter5-VLA18, which is coincident with two pulsars, we compare the continuum flux to the combined fluxes from the timing analysis of Ter5Z and Ter5ae. For the redbacks (Ter5A, Ter5P and Ter5ad), spectral indices from the timing analysis are not available or difficult to interpret due to eclipses (e.g., \citealt{2019ApJ...877..125B}), so we assume a value of $\alpha=-1.5$. We compare these values to the 2.6$\,\GHz$ flux densities from our continuum data (Table \ref{tab:radio}). 

For the 18 previously known MSPs detected in our data, the ratio of the timing to continuum flux densities is consistent with unity, with an RMS scatter of 0.2\,dex (Figure \ref{fig:pul_flux_comp}). Some scatter is expected beyond simple observational uncertainties due to refractive scintillation; for Terzan\,5, at 2\,GHz, the expected RMS variation is 25-30\%  \citep{2019arXiv190708395H}. For 19 of the 20 remaining timed MSPs, the predicted 2.6 GHz flux densities range from $\sim 4$-- 27 $\mu$Jy. Since the 2.6 GHz image has a $5\sigma$ ($3\sigma$) detection floor of $> 32 \mu$Jy ($> 19 \mu$Jy), the non-detection of these fainter MSPs is expected.

\begin{figure}
    \centering
    \includegraphics[width=0.46\textwidth]{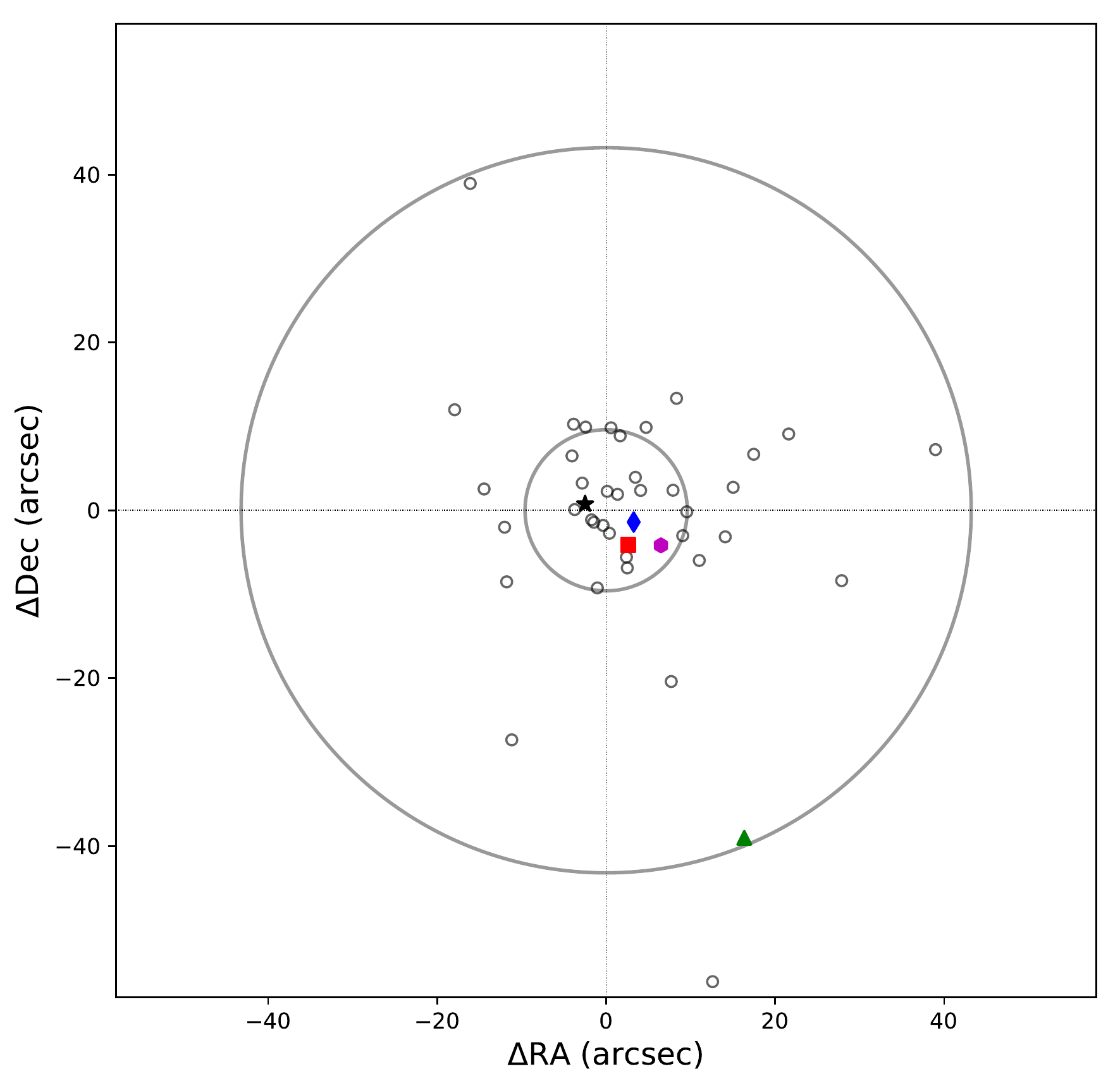}
    \caption{Location of pulsars within Terzan\,5. Open black circles indicate the positions of confirmed MSPs \citep{2005Sci...307..892R,2006Sci...311.1901H, 2017ApJ...845..148P,2018ApJ...855..125C}. Colored symbols indicate new sources within the half-light radius, including candidate pulsars: Ter5-VLA34 (red square), Ter5-VLA38 (blue diamond), Ter5-VLA40 (magenta hexagon); a stellar-mass black hole candidate (Ter5-VLA31; black star); and a possible background source (Ter5-VLA41; green triangle). The core and half-light radii are indicated by the smaller and larger concentric black circles, respectively.}
    \label{fig:pul_dist}
\end{figure}

The one exception is Ter5ad, a redback with a predicted flux density bright enough for us to have detected it in our continuum data: its predicted flux density is $\sim 35$--50 $\mu$Jy at 2.6 GHz. One possibility is that it is slightly fainter than reported in  \citet{2006Sci...311.1901H}, putting it below our detection threshold at 2.6 GHz. Another possibility is that it is eclipsed for some or all of our S band observations. Extrapolating the published ephemerides to our 2012 epoch, an eclipse is not predicted during either of our 2.5 hr blocks; on the other hand, redbacks can show irregular eclipses (e.g., \citealt{2020MNRAS.tmp..561P}). Deeper data obtained over a larger fraction of Ter5ad's orbit could help distinguish between these possibilities. The other previously known redbacks are all detected in our data, though possibly with larger scatter around the 1:1 flux relation, perhaps suggesting they could be partially affected by eclipses as well.

We note that the previous VLA radio continuum imaging of \citet{2000ApJ...536..865F} reported the detection of five sources within the half-light radius of Terzan 5. Four of these match timed MSPs, but one does not: the source that they call N', which is north of the cluster core, and detected only at their lowest observing frequency of 330 MHz. If this source has persistent emission, its non-detection in our data suggests a steep spectrum of $\alpha \lesssim -2.6$. New, high-resolution data at a frequency below 2 GHz would be necessary to confirm the existence and spectral properties of the source.

\begin{figure}[t]
    \centering
    \includegraphics[width=0.46\textwidth]{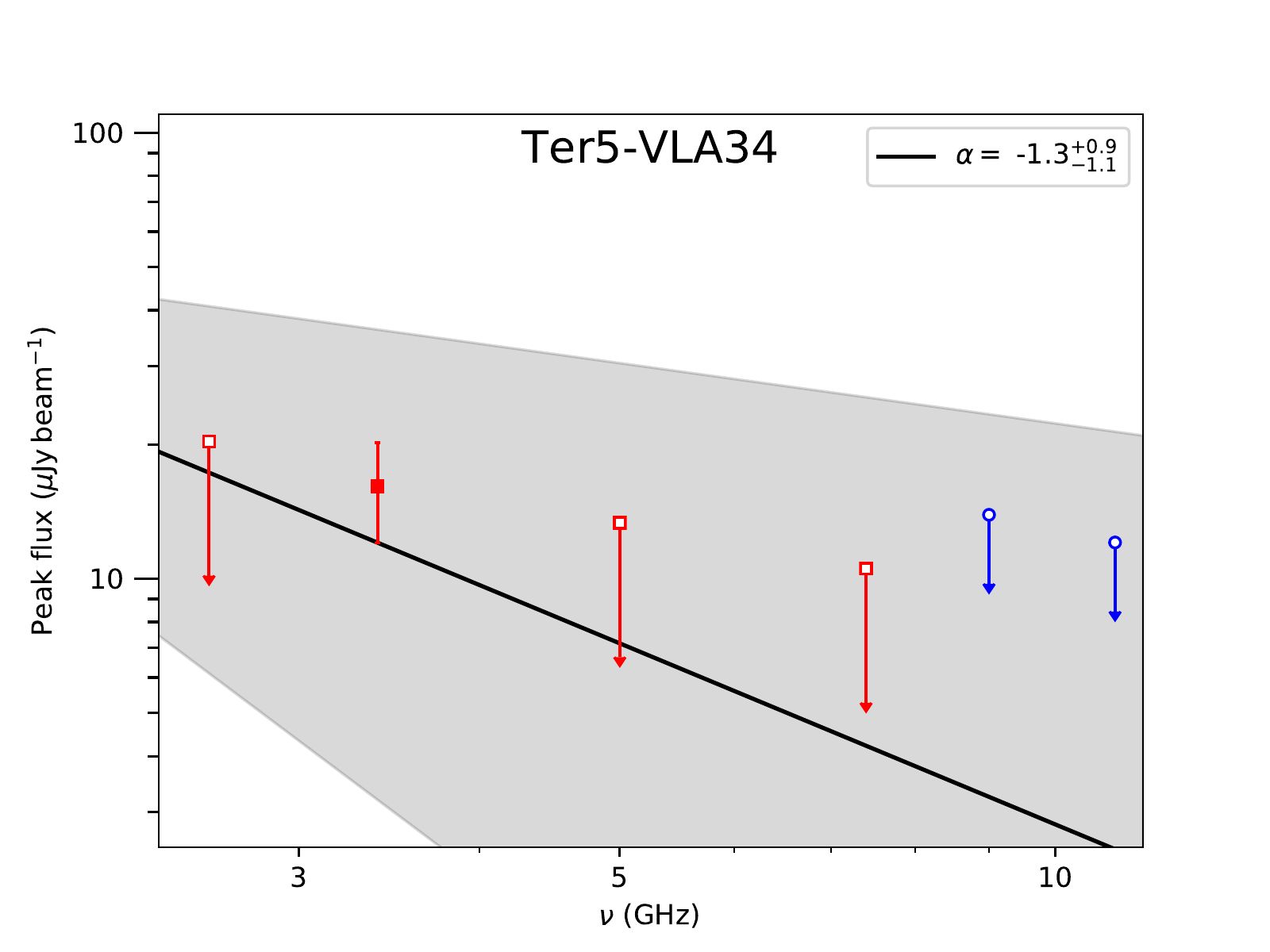}\\
    \includegraphics[width=0.46\textwidth]{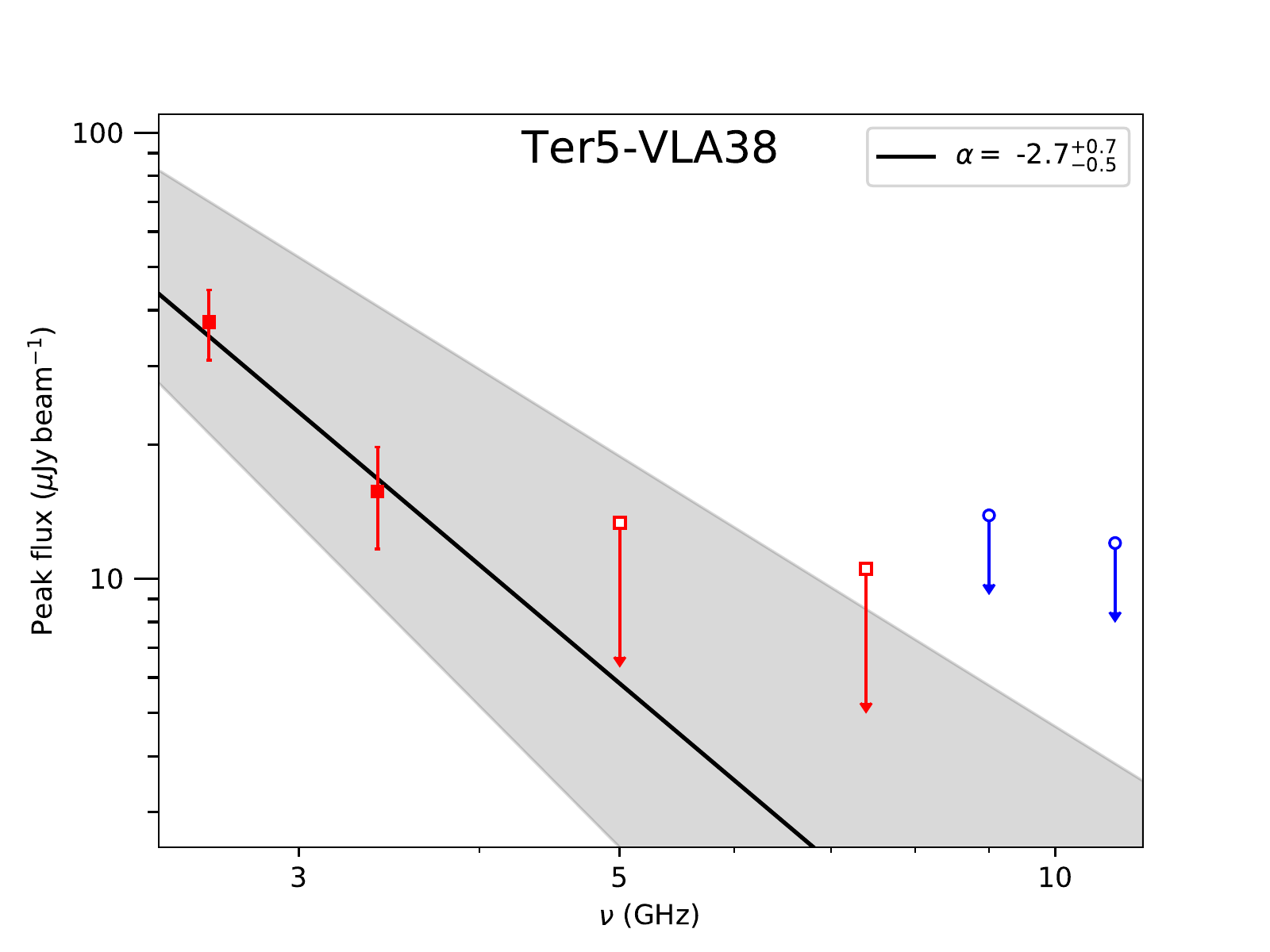}\\
    \includegraphics[width=0.46\textwidth]{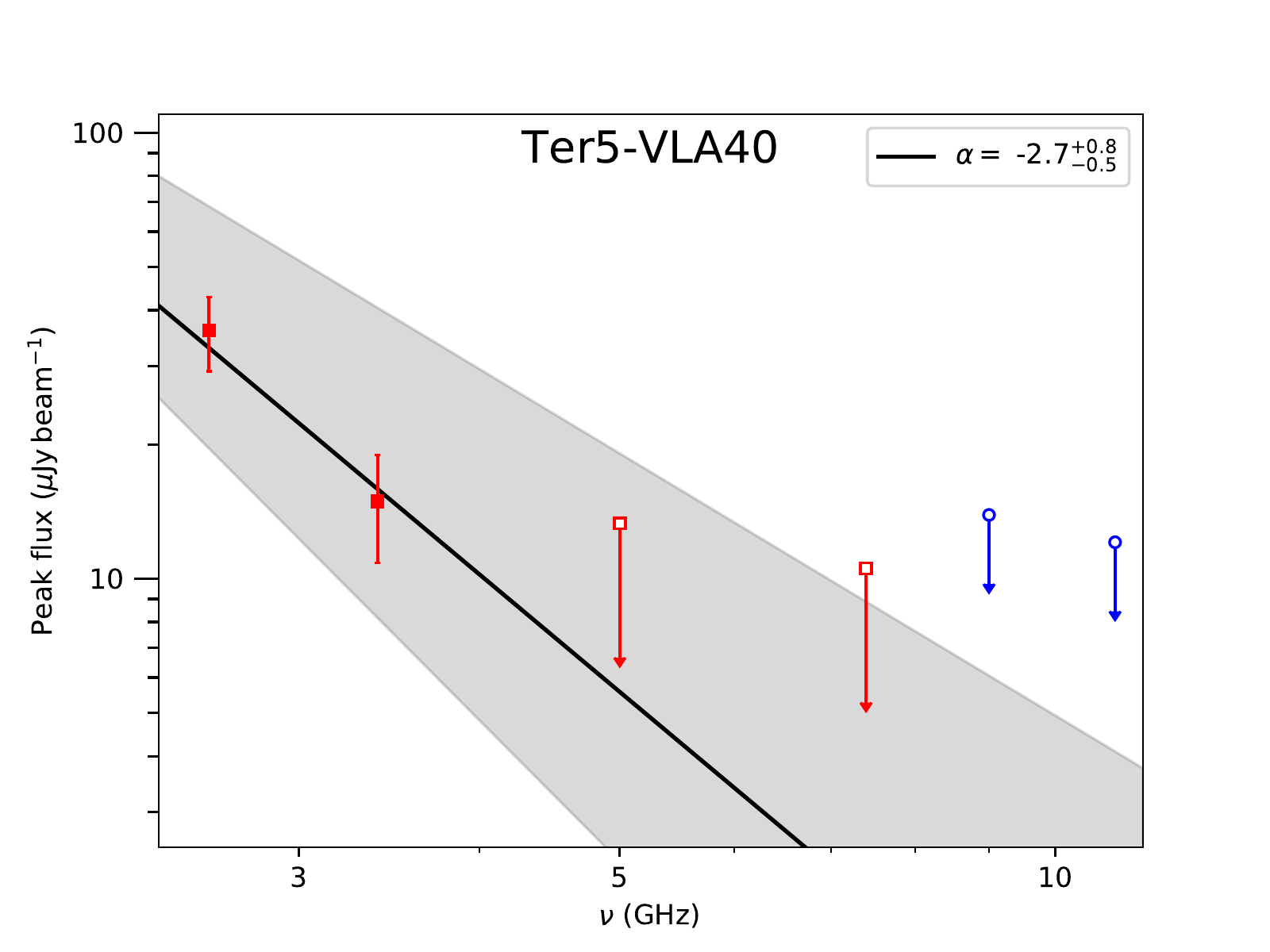}
    \caption{Broadband radio spectra for the three spider MSP candidates. Red squares indicate the quasi-simultaneous 2012 VLA observations at frequencies 2.6, 3.2, 5.0 and 7.4\,GHz. Blue circles represent the 9.0 and 11.0\,GHz VLA observations from 2014. Open data points indicate 3$\sigma$ upper limits. The best-fitting spectral index (black solid line) is calculated using only the quasi-simultaneous 2012 (red) points. The shaded regions indicate the 1$\sigma$ uncertainties on the spectral index.}
    \label{fig:ter5_pulsars}
\end{figure}

\subsection{New redback pulsar candidates} \label{sec:pul_can}

In addition to detecting all of the brighter, non-eclipsing known MSPs, we have discovered three sources that are strong candidates as new MSPs not yet found in timing data. These sources are marked in Figure \ref{fig:pul_dist} and are listed in Table \ref{tab:radio} as Ter5-VLA34, Ter5-VLA38 and Ter5-VLA40. All three sources are located in the mass-segregated cluster core and have associated X-ray sources, and Ter5-VLA38 and Ter5-VLA40 have the steep spectral indices ($\alpha < -2$) characteristic of MSPs (the situation for Ter5-VLA34 is more complicated and discussed in detail later).

Ter5-VLA38 and Ter5-VLA40 notably have 2.6 GHz flux densities of 36--38 $\mu$Jy, brighter than more than half the known MSPs in Terzan 5. These relatively high flux densities mean that the sources should not to have been missed in previous MSP searches due to low S/N alone. Binary MSPs can also be missed if in eclipsing systems (redbacks or black widows) or in exceptionally accelerated binaries. We now discuss these candidates individually and show that there is good evidence all of these new pulsars are redbacks or black widows: eclipsing MSP binaries with non-degenerate companions.

\subsubsection{Terzan\,5 source 38}
There is compelling radio and X-ray evidence that Ter5-VLA38 is a redback. The source is located within the cluster core, 3\barcs3 from the center. It is only detected at low frequencies ($<$ 4 GHz), with a steep best-fitting radio spectral index of $\alpha=-2.7_{-0.5}^{+0.7}$ (Figure \ref{fig:ter5_pulsars}). We note that Ter5-VLA38 was also clearly detected at 3.2 GHz in 2014 (Figure \ref{fig:ter5_high_freq}). Ter5-VLA38 also has a coincident X-ray detection. When fit with an absorbed power-law model, we find a best-fitting photon index of $\Gamma=1.2\pm0.2$ and unabsorbed 0.5--10\,keV luminosity, $L_{\rm X}=\left(2.4\pm0.1\right)\times10^{32}\,\ergs$. 

\begin{figure}
    \centering
    \includegraphics[width=0.46\textwidth]{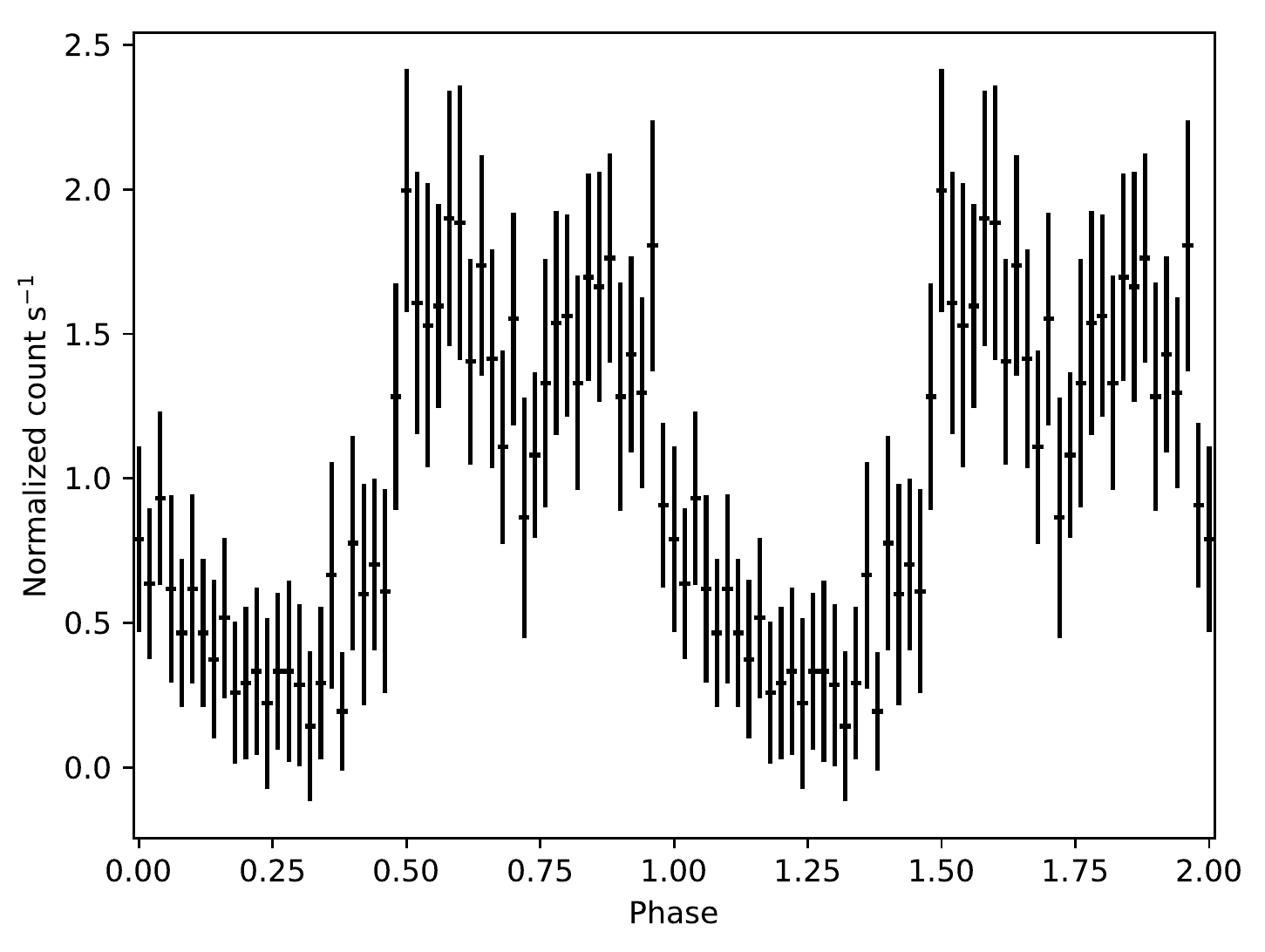}\\
    \includegraphics[width=0.46\textwidth]{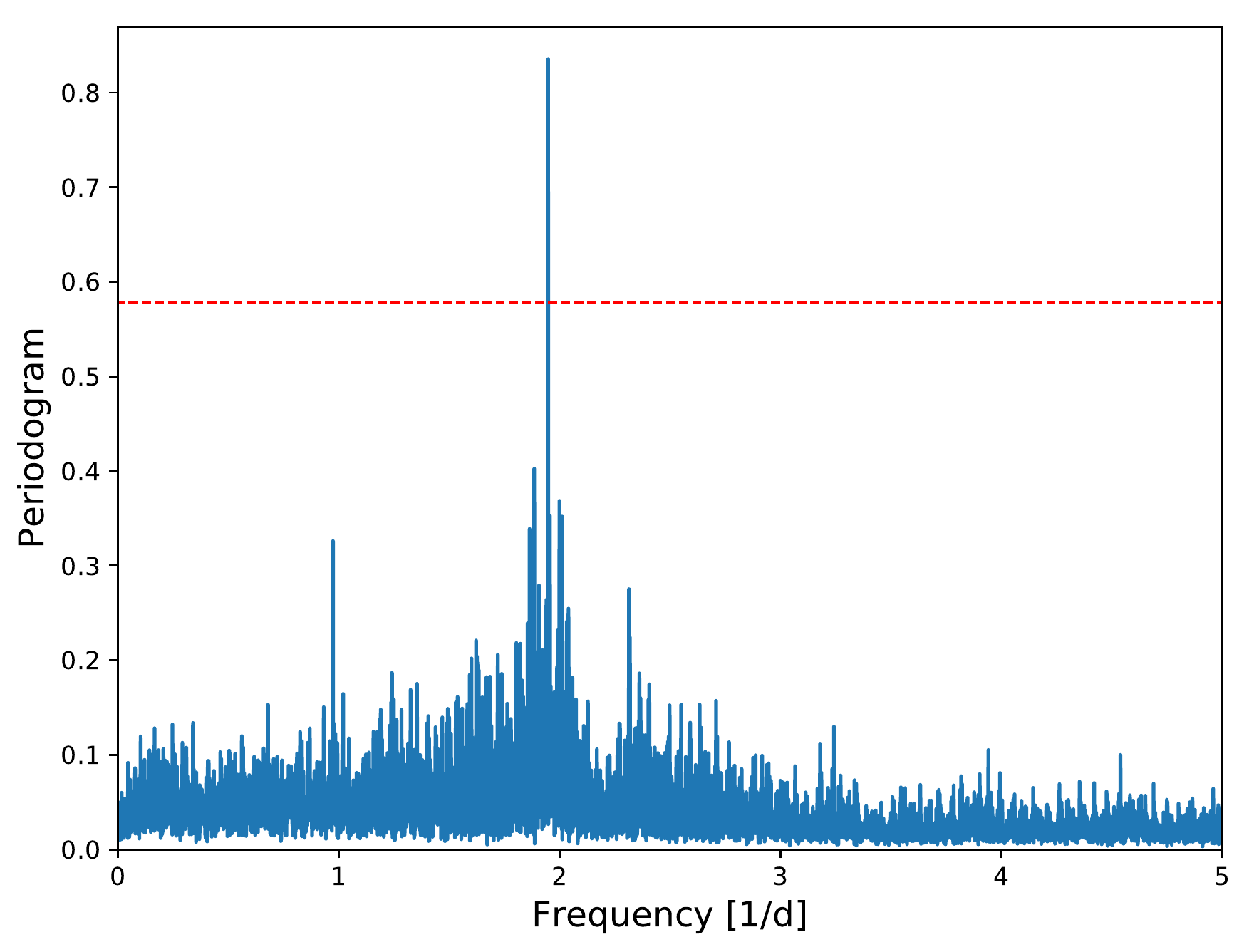}
    \caption{Top: Folded background-subtracted \textit{Chandra}/ACIS-S $0.3$--7\,keV light curve of Ter5-VLA38. The best-fitting period is 12.32 hr. The absolute phase is arbitrarily set to have X-ray maximum around $\phi=0.75$.
    Bottom: power spectrum for the X-ray light curve of Ter5-VLA38. The red dashed line indicates $3\sigma$ significance. The best-fitting period 0.51\,d has a significance of $>4\sigma$. The periodic X-ray emission on timescales of $\lesssim1$\,d, along with the ``double-peaked'' light curve are both consistent with Ter5-VLA38 being a redback source.}
    \label{fig:lc_src38}
\end{figure}

The X-ray source is highly variable. We test for periodicity using the python package \texttt{P4J}'s implementation of PDM \citep{1978ApJ...224..953S}. The PDM is ideal for irregularly sampled data and non-sinusoidal profiles, such as the Terzan\,5 \textit{Chandra} observations used here. We find a best-fitting period of $0.5133^{+0.0001}_{-0.0003}$\,d, at a significance greater than $4\sigma$ (Figure \ref{fig:lc_src38}). The phased light curve shows a clear orbital modulation along with a striking double-peaked maximum, very similar to that observed in known redbacks such as PSR J2129-–0429 \citep{2018ApJ...861...89A}. In this system the basic orbital modulation is explained as a Doppler-boosted intrabinary shock observed close to edge on. The double-peaked maximum at its observed phase
is more difficult to explain, and could be due to an unexpectedly high momentum flux from the companion wind causing the shock to wrap around the pulsar \citep{2016ApJ...828....7R,
2018ApJ...869..120W}. Since we have no absolute phase information available for Ter5-VLA38 we cannot say whether the shock in this system appears to wrap around the pulsar or the secondary, which might be easier to explain.

In any case, Ter5-VLA38's orbital period, X-ray light curve, X-ray photon index and luminosity (see Figure \ref{fig:pulsar_gamma_vs_lx} and discussion below), and radio spectral index are all consistent with known redbacks, making it an exceptionally strong candidate redback MSP.


\subsubsection{Terzan\,5 source 40}
Ter5-VLA40 is located within the core, and is detected at both 2.6 and 3.4 GHz. Ter5-VLA40 has a steep radio spectral index of $\alpha=-2.7_{-0.5}^{+0.8}$, characteristic of MSPs. Ter5-VLA40 also has an associated X-ray counterpart, with photon index $\Gamma=1.3_{-0.7}^{+1.0}$ and unabsorbed 0.5--10\,keV luminosity of $L_{\rm X}=\left(1.5_{-0.4}^{+0.9}\right)\times10^{31}\,\ergs$, as well as evidence for X-ray variability (though the source is too dim for a useful X-ray light curve). While the best-fitting photon index for Ter5-VLA40 is not as well-constrained as for Ter5-VLA38 or Ter5-VLA34 (see below), it is still consistent with the properties of known eclipsing spider MSPs.

\subsubsection{Terzan\,5 source 34}
Ter5-VLA34 is located within the core of Terzan\,5, $4\barcs8$ south-west of the cluster center. This source is marginally detected in our data; it was only detected in the 2012 observations at 3.4\,GHz at a S/N $\approx 4$. Nonetheless, it has an associated X-ray source, giving confidence in the radio detection.
Its radio spectral index is poorly constrained, $\alpha= -1.3_{-1.1}^{+0.9}$ (Figure \ref{fig:ter5_pulsars}), making it difficult to differentiate between a steep-spectrum pulsar and a flatter-spectrum low-mass X-ray binary based on radio properties alone. Its 3.4\,GHz flux density is comparable to the other new MSP candidates Ter5-VLA38 and Ter5-VLA40, so if it is an MSP as well, the non-detection at 2.6 GHz could be due to a flatter spectral index or to random flux uncertainties.

\begin{figure*}
    \centering
    \includegraphics[width=0.8\textwidth]{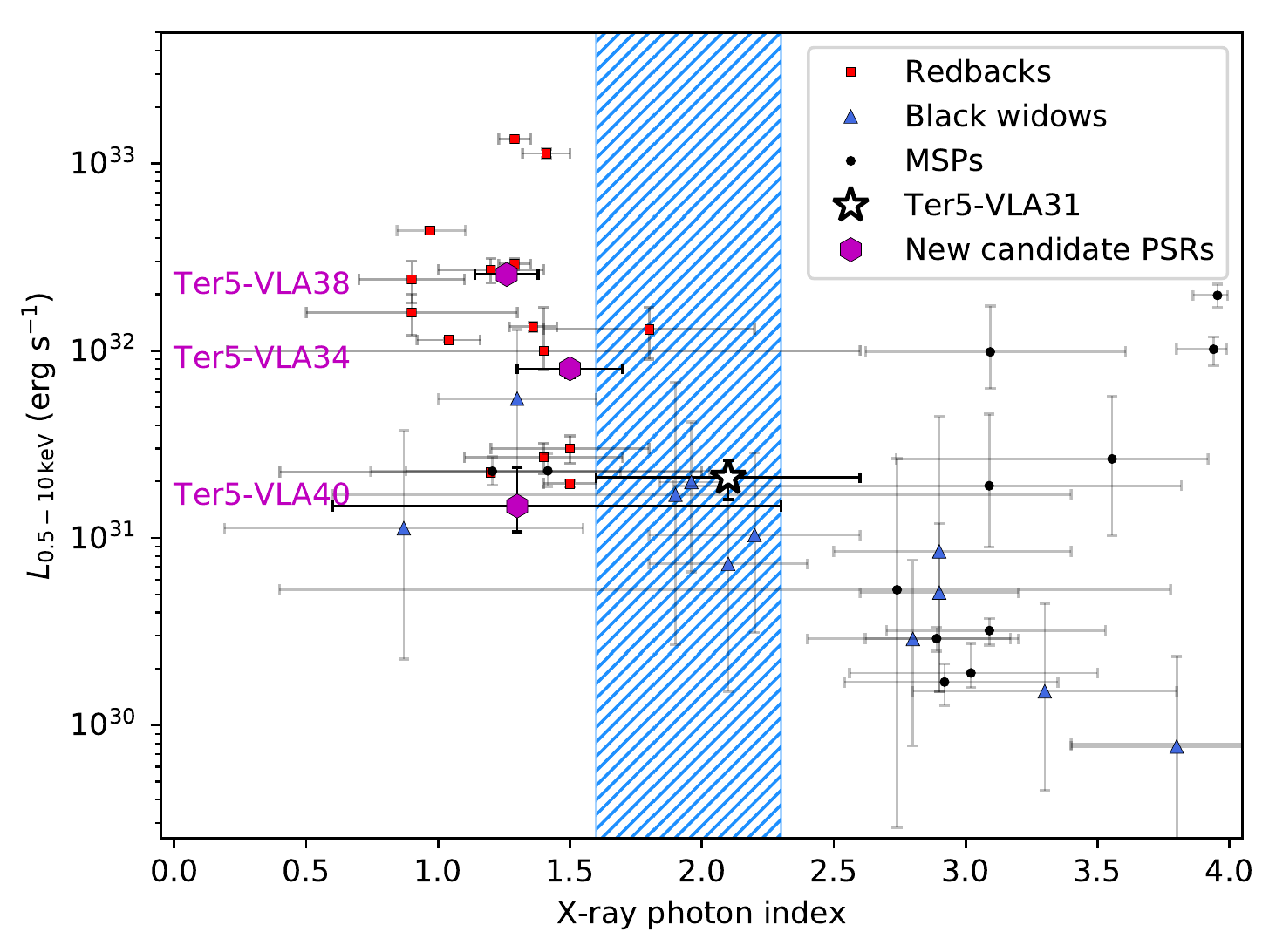}
    \caption{X-ray luminosity (0.5--10 keV) vs.\ photon index for a selection of compact objects. Red squares represent known redbacks \citep{2014ApJ...795...72L,2015arXiv150207208R,2015ApJ...814...88S,2017ApJ...838..124H,2017MNRAS.465.4602L,2018ApJ...866...71C,2018MNRAS.473L..50L,2018ApJ...863..194L,2018ApJ...861...89A,2018ApJ...866...83S,Strader19}, blue triangles represent known black widows \citep{2018ApJ...864...23L}, while black circles represent known, non-eclipsing MSPs in Terzan\,5 \citep{goose} and 47 Tuc \citep{2017MNRAS.472.3706B}. The magenta hexagons represent our candidate pulsars that have an X-ray counterpart (Ter5-VLA34, Ter5-VLA38 and Ter5-VLA40); our candidate pulsars occupy a similar parameter space to other eclipsing MSPs. Additionally, we also include the parameter space containing $>90\%$ of quiescent stellar-mass black holes (from \citealt{2013ApJ...773...59P}; hatched blue region) and our candidate stellar-mass black hole (Ter5-VLA31; open black star) for comparison.}
    \label{fig:pulsar_gamma_vs_lx}
\end{figure*}

The X-ray properties give clues to the true nature of the source: the counterpart is significantly variable, with an 
average unabsorbed 0.5--10\,keV luminosity of $\left(8.0\pm0.8\right)\times10^{31}\ \ergs$ and a well-constrained photon index of $\Gamma= 1.5_{-0.2}^{+0.2}$ (Table \ref{tab:xray}).
This X-ray luminosity would be consistent with the range of quiescent low-mass X-ray binaries with neutron stars in globular clusters \citep[e.g.,][]{2003ApJ...598..501H}. However, no quiescent neutron star (with $L_X < 10^{33}$\,\ergs) has ever been detected in the radio, except for the transitional MSPs discussed below, despite deep searches (e.g., \citealt{2017MNRAS.470..324T}).
We also note that the photon index is harder than the majority of quiescent accreting black holes \citep{2013ApJ...773...59P}. Instead, Ter5-VLA34's joint X-ray/radio properties are most consistent with a spider MSP, where the variable X-rays arise from an intrabinary shock (Figure \ref{fig:pulsar_gamma_vs_lx}). We use a Lomb-Scargle analysis to test for periodicity in the X-ray emission on the timescale of $\lesssim1$\,d. However, no significant periodic signal was identified. If the source is periodic, as expected for black widows or redbacks, we currently lack sufficient S/N for a significant detection. While the current evidence suggests that Ter5-VLA34 is in fact a spider MSP binary, future deep radio observations would be valuable to better constrain the radio spectral index of this likely MSP.

\subsection{Comparison to known pulsar population}

Here we compare the X-ray and radio properties of our new sources to those of known pulsars. 

Figure \ref{fig:pulsar_gamma_vs_lx} plots X-ray luminosity against photon index for our new discoveries and for known populations of compact objects. It is clear that the X-ray luminosities and photon indices of our MSP candidates are consistent with known spider MSPs. In particular, the high X-ray luminosities and hard photon indices of Ter5-VLA38 and Ter5-VLA34 make them excellent redback candidates. In the case of Ter5-VLA38 this classification is bolstered by
its X-ray light curve and long (12.32 hr) orbital period.

For Ter5-VLA40, the lower X-ray luminosity and poorly determined photon index makes its nature less certain: it falls into an area of parameter space occupied by both redbacks and black widows. The detection of a pulsar (or an optical/IR companion) associated with this sources may be necessary to improve this classification.

All three sources also show strong evidence of variability, which is a feature of field redbacks \citep{2014AN....335..313R, 2015ApJ...801L..27H} as well as the known redbacks Ter5A and Ter5P in Terzan 5 (Table \ref{tab:xray}).

In Figure \ref{fig:pul_flux_comp}, we compared the associated fluxes determined via timing and continuum analyses for each source that is detected in both data sets. Two of our new candidate pulsars, Ter5-VLA38 and Ter5-VLA40, have radio flux densities higher than nearly half of the pulsars previously detected by timing surveys, and the third candidate (Ter5-VLA34) is only slightly fainter. Thus, we would expect all of our candidate redbacks to have been detected in previous timing searches. They may have been missed \textit{(i)} if they are redback or black widow binary MSPs, where the radio pulses are eclipsed for much or all of the orbit by material blown off the companion star or; \textit{(ii)} if they are in highly accelerated binaries, e.g., short-period MSP--black hole binaries that are difficult to find in standard acceleration searches. The multi-wavelength data for Ter5-VLA34, Ter5-VLA38, and Ter5-VLA40 are all more consistent with explanation \textit{(i)}.

A confirmation of these classifications would require either the detection of MSPs in these candidate binaries or of optical/IR counterparts. The latter is challenging due to the high foreground reddening toward Terzan 5 ($E(B-V) \sim 2.4$), though a variability study in the near-IR could be beneficial. It may be that a reanalysis of older pulsar timing data or new observations could reveal MSPs in these binaries.

\subsection{New stellar-mass black hole candidate} \label{sec:bh_can}

We have identified one accreting stellar-mass black hole candidate: Ter5-VLA31. Ter5-VLA31 is distinguished from the candidate pulsars by its well-determined,
flat radio spectral index of $\alpha=-0.2\pm0.2$. Its position within the core of Terzan\,5 ($r\approx2\farcs4$) suggests it is very unlikely to be an unassociated background source; in fact, Ter5-VLA31 is the only flat-spectrum source within the cluster half-light radius ($43\farcs2$; Figure \ref{fig:radvspec}). 

\begin{figure}
    \centering
    \includegraphics[width=0.46\textwidth]{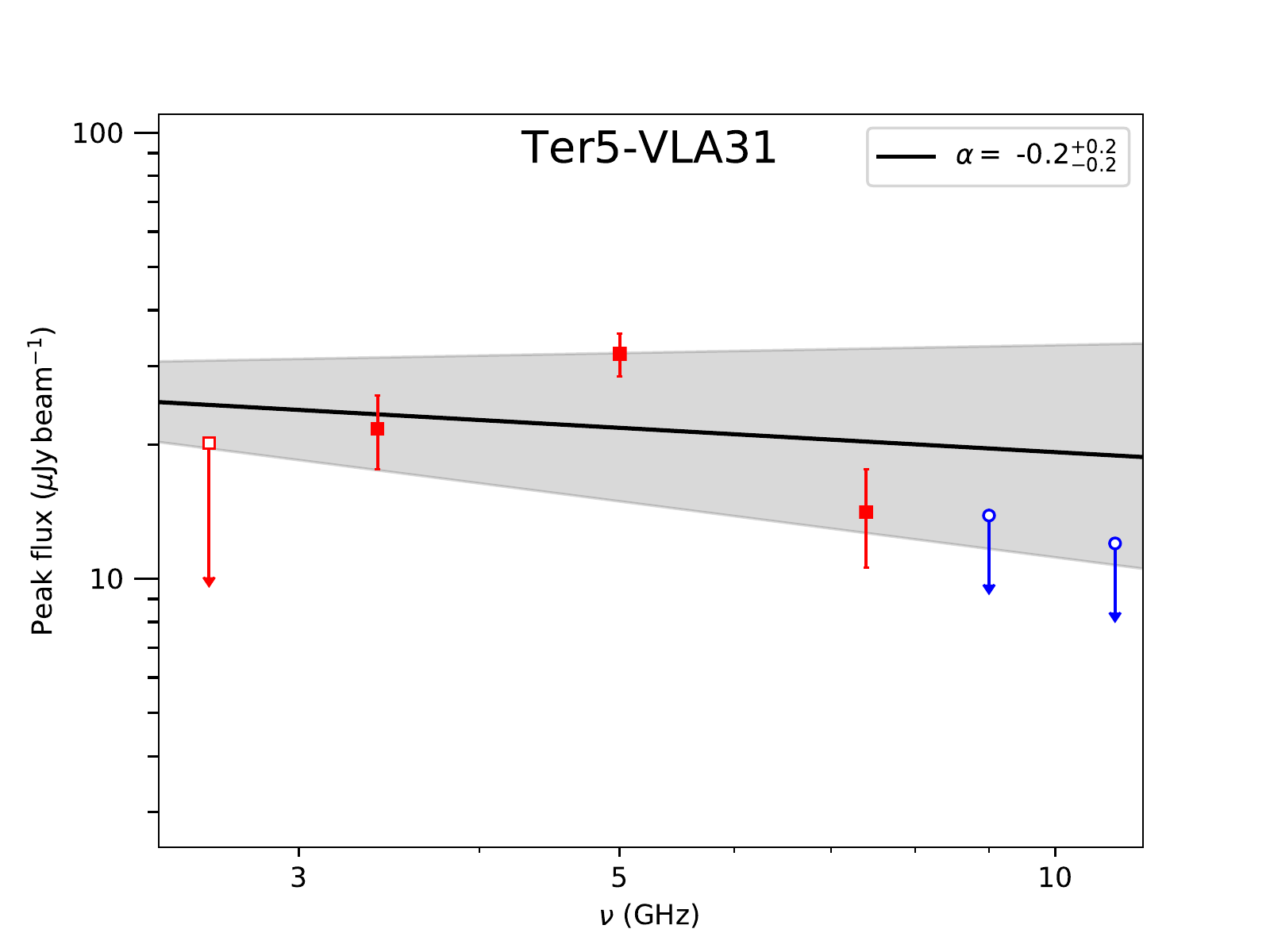}
    \caption{Radio spectrum of the candidate stellar-mass black hole candidate Ter5-VLA31. The symbols are as in Figure \ref{fig:ter5_pulsars}.
    We identify this source as a candidate black hole due to its flat spectral index ($\alpha=-0.2\pm0.2$), X-ray luminosity and spectrum, and close proximity to the cluster core.}
    \label{fig:ter5_bh}
\end{figure}

Ter5-VLA31 is detected at 3.4, 5.0 and 7.4\,GHz in the 2012 quasi-simultaneous VLA observations (Figure \ref{fig:ter5_bh}). It is not detected in the 2012 2.6\,GHz image, but our 3$\sigma$ upper limit is consistent with the flux density measurement at 3.4 GHz, and the 2.6\,GHz image has the lowest resolution and hence is crowded at the location of the source. Ter5-VLA31 is also detected at 3.2 GHz in the 2014 image at an approximately equivalent flux density. While we do not use this measurement in our analysis below, these data lend general credence to the existence and properties of this source.

\begin{figure*}
    \centering
    \includegraphics[width=0.8\textwidth]{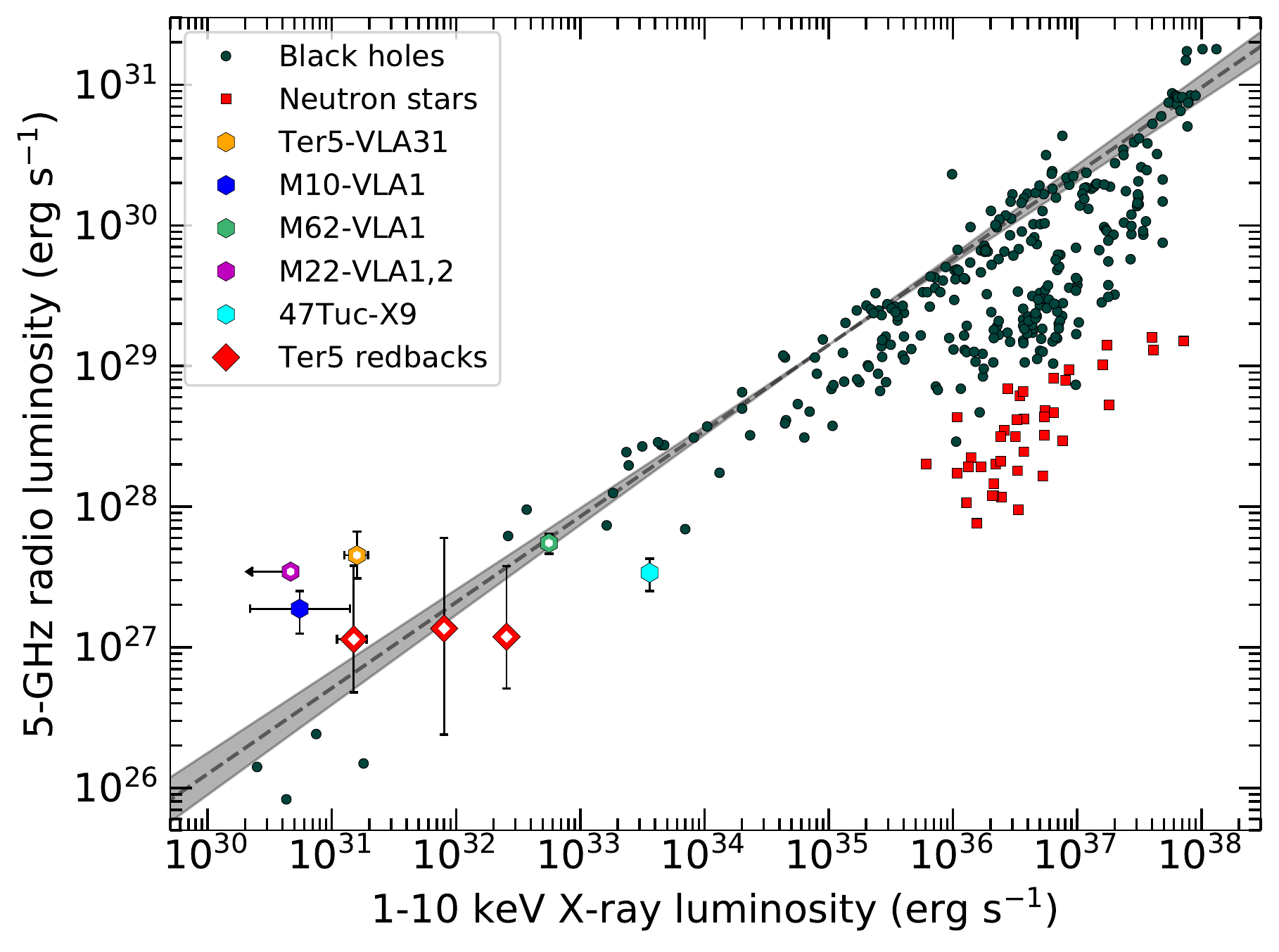}
    \caption{Radio luminosity vs.\ X-ray luminosity of low-mass X-ray binaries and sources discussed in this paper. Black circles represent quiescent/hard-state stellar-mass black holes \citep{Tetarenko16}. Red squares represent accreting neutron stars from the compilation of \citet{arash_bahramian_2018_1252036}. The orange hexagon represents the Terzan\,5 black hole candidate (Ter5-VLA31), while the remaining colored hexagons represent other MAVERIC black hole candidates in globular clusters; M62-VLA1 \citep{2013ApJ...777...69C}, 47\,Tuc-X9 \citep{2015MNRAS.453.3918M}, M10-VLA1 \citep{2018ApJ...855...55S} and M22-VLA1 and M22-VLA2 (these sources largely overlap and thus are represented as a single point; \citealt{2012Natur.490...71S}). In contrast, the red diamonds represent the three candidate redbacks in Terzan\,5, for which we do not find evidence of accretion: Ter5-VLA40, Ter5-VLA34 and Ter5-VLA38 (in ascending X-ray luminosity). Open markers indicate non-simultaneous X-ray and radio observations. The dashed line shows the best-fit radio/X-ray correlation for low/hard state stellar-mass black holes \citep{Gallo14}, with the shaded region outlining the 1$\sigma$ error.}
    \label{fig:lrlx}
\end{figure*}

One oddity found during our source characterization is that the 5.0\,GHz counterpart appears to be offset (by $\sim1\farcs3$), from the mean weighted position of the 3.2, 3.4 and 7.4\,GHz detections (the average 1$\sigma$ positional uncertainties for each measurement are of the order $0\farcs1$--$0\farcs2$ in both RA and Dec). While it is possible that the 5.0\,GHz detection constitutes a different source, we think this is unlikely because \textit{(i)} we do not detect anything at the location of the 5.0\,GHz source in any other radio frequency or in X-rays, \textit{(ii)} the flux density of the source at 5.0 GHz is generally consistent with that expected based on the lower and higher frequency data. As another test, we also divided the 5.0 GHz data into five epochs of equal duration, finding the source to be coincident with the 3.2/3.4/7.4 GHz mean-weighted position of Ter5-VLA31 in two of these sub-epochs. Overall, our 5.0 and 7.4 GHz data has relatively poor $uv$ coverage, and better high-resolution images would be valuable in understanding this source. The current data are most consistent with the idea that these flux densities all represent a single source, and we proceed on that basis.

An X-ray source is detected at the radio position of Ter5-VLA31. The X-ray counterpart is consistent with being moderately soft; when fit with a power-law, we find a photon index of $\Gamma=2.1\pm0.5$. We measure a time-averaged unabsorbed 0.5--10\,keV luminosity of $L_{0.5-10\,\kev}=\left(2.1\pm0.5\right)\times10^{31}$ erg s$^{-1}$ ($L_{1-10\,\kev}=\left(1.6_{-0.3}^{+0.4}\right)\times10^{31}$ erg s$^{-1}$). In Figure \ref{fig:lrlx}, we compare the X-ray and radio luminosities of Ter5-VLA31 with other compact objects, keeping in mind that the radio/X-ray relation for black hole X-ray binaries is poorly constrained at low luminosities. Ter5-VLA31 falls significantly above the black hole correlation, implying that it is radio bright for the measured X-ray luminosity, though unsurprising as it was selected for its radio brightness.

It is important to note that the black hole correlation is constructed from simultaneous measurements, and since we know quiescent LMXBs show short and long-term variations in both their X-ray and radio fluxes (e.g., \citealt{2019ApJ...874...13P}), in principle the lack of simultaneous measurements makes the interpretation more challenging. However, Ter5-VLA31 does not show significant X-ray variability over short or long timescales \citep{goose}.

While the current radio and X-ray properties of Ter5-VLA31 are suggestive of an accreting stellar-mass black hole, other scenarios may also be viable (see \citealt{2012Natur.490...71S,2015MNRAS.453.3918M,2018ApJ...855...55S} for more comprehensive discussions on alternative classifications for candidate stellar-mass black holes). For instance, radio flares have been detected from accreting white dwarfs (refer to \citealt{2015MNRAS.453.3918M} for a detailed analysis of radio emission from white dwarfs). However, this emission tends to be at least an order of magnitude lower in luminosity than what we see from Ter5-VLA31. Alternatively, the source could be an accreting neutron star, though we would expect several orders-of-magnitude higher X-ray luminosity for our measured radio luminosity (Figure \ref{fig:lrlx}). We note two important caveats regarding this interpretation: firstly, X-ray-bright accreting neutron stars are being discovered at luminosities comparable to black holes (e.g., \citealt{2020MNRAS.492.1091G, 2020MNRAS.493.1318V}), and; secondly, our X-ray and radio observations are not simultaneous, but we are instead relying on a time-averaged (over $\sim15\,$yr) X-ray luminosity. However, as mentioned above, Ter5-VLA31 does not appear to be significantly X-ray variable (Table \ref{tab:xray}). The lack of clear variability in the X-ray emission is also evidence against Ter5-VLA31 being a transitional MSP (sources that are also known to produce flat-spectrum radio emission while in their accretion-powered state; e.g., \citealt{2015ApJ...809...13D}). Additionally, Ter5-VLA31 has an X-ray luminosity below the typical range displayed by transitional MSPs in their accreting states, $10^{32}-10^{34}\,\ergs$ \citep{2013A&A...550A..89D, 2013Natur.501..517P, 2014ApJ...781L...3P}. 
Overall, a classification as a quiescent stellar-mass black hole best fits the existing data, but simultaneous high-resolution radio and X-ray observations could help constrain alternative scenarios.

\subsection{Terzan\,5 source 41: A likely background source}
Ter5-VLA41 is located at the edge of the half-light radius, $42\arcs$ south-east of the cluster center. It is only detected at 2.6\,GHz, so we can only place an upper limit on the spectral index, $\alpha<0.1$. Additionally, it has no X-ray counterpart, with an X-ray upper limit of $\lesssim 1.2\times 10^{30}$ erg s$^{-1}$. Because of this, and the fact that it is located far from the core, we suggest Ter5-VLA41 to be a low-probability compact binary candidate. As three known MSPs in Terzan 5 have been discovered at least as far from the center (Figure \ref{fig:pul_dist}), the possibility that this source is indeed a pulsar cannot be definitively ruled out, though it is not all that faint and hence could in principle
have been detected in previous timing observations.

\section{Conclusions} \label{sec:conclusion}

We have performed deep, high-resolution 2--8\,GHz VLA observations centered on the Galactic globular cluster Terzan\,5. A total of 43 radio continuum sources have been identified. Of these, 24 sources are likely associated with the cluster, with the remainder as probable background sources. Combining our radio data with a comprehensive X-ray catalog, we have discovered:

\begin{itemize}
\item Three new steep-spectrum radio sources (Ter5-VLA34, Ter5-VLA38, and Ter5-VLA40) with variable X-ray counterparts and best-fitting power-law photon indices $\Gamma=1.3-1.5$. 
For Ter5-VLA38 we find an X-ray orbital period of 12.32 hr and a well-defined X-ray light curve with a double-peaked maximum.
The X-ray and radio properties of Ter5-VLA38 and Ter5-VLA34 strongly suggest they are redback MSPs; for Ter5-VLA40 the evidence is consistent with either a redback or black widow classification. The continuum flux densities of these sources are well above many previously discovered MSPs in Terzan 5, suggesting they were missed due to radio eclipses.

\item One flat-spectrum radio source close to the center of Terzan\,5, Ter5-VLA31. This object has an X-ray counterpart with $\Gamma=2.1\pm0.5$ and $L_{\mathrm X}=(2.1\pm0.5)\times10^{31}\,$\ergs. The X-ray and radio properties of Ter5-VLA31 are more consistent with that of quiescent stellar-mass black holes than other classes of compact binary. Its position on the radio/X-ray correlation for low-mass X-ray binaries is near that of other globular cluster black hole candidates.

\item 17 radio continuum counterparts to MSPs previously discovered through rigorous single-dish timing surveys. Our continuum flux measurements are in general agreement with previous pulsed flux estimates. The 20 timed MSPs we did not detect are likely below our continuum detection limit ($\sim20\,\mu\jansky\,\beam$ at 3\,GHz; Figure \ref{fig:pul_flux_comp})

\end{itemize}

The discovery of three new, relatively bright spider MSPs  highlights the ability of deep radio continuum imaging to complement pulsar time-domain searches to find unusual pulsar binaries. If the redbacks follow a similar radio luminosity function to the underlying MSPs, then new high-resolution radio data that can go a factor of 2--3 deeper in sensitivity could be expected to discover 4-10 more eclipsing MSPs. Higher angular resolution could give an additional boost, as our main low-frequency data were obtained in BnA configuration rather than the most extended VLA A configuration. Finally, data taken over several epochs could allow radio eclipses to be discovered, offering an alternative path to confirming new redbacks.

\acknowledgments
We thanks the anonymous referee for their helpful feedback. RU and LC are grateful for support from {\it Chandra} award GO7-18032A and {\it HST} award HST-GO-14351. We acknowledge support from NSF grants AST-1308124 and AST-1514763. JS acknowledges support from the Packard Foundation. SMR is a CIFAR Fellow and is supported by the NSF Physics Frontiers Center award 1430284. COH and GRS acknowledge NSERC Discovery Grants RGPIN-2016-04602 and RGPIN-2016-06569 respectively; COH also acknowledges a Discovery Accelerator Supplement. JCAM-J is the recipient of an Australian Research Council Future Fellowship (FT140101082), funded by the Australian government. The National Radio Astronomy Observatory is a facility of the National Science Foundation operated under cooperative agreement by Associated Universities, Inc.  We acknowledge computational support from HPCC (High Performance Computing Center), Michigan State University.

\bibliography{references}

\clearpage
\newpage
\startlongtable
\tabletypesize{\tiny}
\centerwidetable
\tablewidth{0pt} 
\rotate
\begin{deluxetable}{lccccccccccccc}
\tablecaption{Radio sources in Terzan\,5.\label{tab:radio}}
\tablehead{
\colhead{\#} & \colhead{RA (ICRS)} & \colhead{$\sigma_{\mathrm RA}$} & \colhead{Dec (ICRS)} & \colhead{$\sigma_{\mathrm Dec}$} & \colhead{Rad.} & \colhead{Alt ID}\tablenotemark{a} & \colhead{$S_{2.6\,GHz}$} & \colhead{$S_{3.4\,GHz}$} & \colhead{$S_{5.0\,GHz}$} & \colhead{$S_{7.4\,GHz}$} & \colhead{$S_{9.0\,GHz}$} & \colhead{$S_{11.0\,GHz}$} & \colhead{$\alpha$} \\
\colhead{} & \colhead{(h:m:s)} & \colhead{(\arcsec)} & \colhead{($^{\circ}$:$^{\prime}$:$^{\prime\prime}$)} & \colhead{(\arcsec)} &  \colhead{(\arcmin)} & \colhead{} & \colhead{(\ujy)} & \colhead{(\ujy)} & \colhead{(\ujy)} & \colhead{(\ujy)} & \colhead{(\ujy)} & \colhead{(\ujy)} & \colhead{}
}
\startdata
1&17:48:02.249&0.05&-24:46:37.65&0.10&0.60&Ter5A &698$\pm$7&307$\pm$4&95$\pm$3&33$\pm$3&$<$14&$<$13&$-3.03_{-0.04}^{+0.04}$\\
2&17:47:50.920&0.05&-24:44:51.09&0.10&3.68&&327$\pm$7&265$\pm$5&223$\pm$5&190$\pm$8&236$\pm$21&&$-0.54_{-0.04}^{+0.04}$\\
3&17:48:06.404&0.05&-24:45:04.94&0.10&1.70&&287$\pm$6&224$\pm$4&174$\pm$4&102$\pm$3&105$\pm$5&92$\pm$6&$-0.92_{-0.03}^{+0.03}$\\
4&17:48:10.920&0.05&-24:45:40.27&0.10&1.75&&192$\pm$6&210$\pm$4&231$\pm$4&189$\pm$3&102$\pm$6&113$\pm$6&$-0.06_{-0.03}^{+0.03}$\\
5&17:48:05.034&0.05&-24:46:41.29&0.10&0.07&Ter5P &224$\pm$7&167$\pm$4&44$\pm$4&$<$11&$<$14&$<$12&$-2.28_{-0.08}^{+0.07}$\\
6&17:47:55.642&0.05&-24:44:58.81&0.10&2.73&&173$\pm$6&147$\pm$4&99$\pm$4&52$\pm$5&$<$29&$<$47&$-1.03_{-0.06}^{+0.06}$\\
7&17:48:04.534&0.05&-24:46:34.70&0.10&0.18&Ter5C  &275$\pm$7&147$\pm$4&46$\pm$4&22$\pm$4&$<$14&$<$12&$-2.56_{-0.09}^{+0.08}$\\
8&17:48:09.898&0.05&-24:48:57.84&0.10&2.50&&108$\pm$6&95$\pm$4&85$\pm$4&53$\pm$5&$<$30&$<$49&$-0.58_{-0.08}^{+0.08}$\\
9&17:48:05.102&0.05&-24:46:34.49&0.10&0.18&Ter5V  &93$\pm$7&66$\pm$4&24$\pm$5&$<$11&$<$14&$<$12&$-2.01_{-0.20}^{+0.18}$\\
10&17:48:16.813&0.05&-24:46:41.82&0.10&2.72&&60$\pm$6&74$\pm$4&67$\pm$4&54$\pm$5&88$\pm$10&$<$54&$-0.19_{-0.10}^{+0.10}$\\
11&17:47:53.156&0.05&-24:46:51.81&0.10&2.66&&71$\pm$6&60$\pm$4&41$\pm$4&30$\pm$7&$<$30&$<$49&$-0.85_{-0.17}^{+0.16}$\\
12&17:47:57.134&0.06&-24:45:20.90&0.12&2.24&&76$\pm$6&55$\pm$4&43$\pm$4&20$\pm$5&$<$22&$<$28&$-1.04_{-0.16}^{+0.16}$\\
13&17:48:08.883&0.16&-24:47:34.68&0.14&1.24&&54$\pm$6&48$\pm$4&39$\pm$4&$<$12&$<$16&$<$17&$-1.02_{-0.17}^{+0.16}$\\
14&17:48:01.732&0.05&-24:46:28.09&0.10&0.76&&41$\pm$7&46$\pm$4&43$\pm$3&34$\pm$3&41$\pm$4&24$\pm$4&$-0.27_{-0.12}^{+0.12}$\\
15&17:48:08.066&0.05&-24:46:01.15&0.10&1.03&&21$\pm$7&45$\pm$4&27$\pm$4&21$\pm$4&$<$15&$<$14&$-0.52_{-0.19}^{+0.19}$\\
16&17:47:49.269&0.16&-24:46:19.62&0.14&3.56&&38$\pm$6&39$\pm$5&$<$19&$<$34& & &$-1.1_{-0.5}^{+0.4}$\\
17&17:48:04.622&0.05&-24:46:40.77&0.10&0.08&Ter5M  &58$\pm$7&37$\pm$4&21$\pm$4&$<$11&$<$14&$<$12&$-1.8_{-0.3}^{+0.3}$\\
18&17:48:04.957&0.05&-24:46:45.93&0.10&0.03&Ter5Z  &66$\pm$7&37$\pm$4&$<$13&$<$11&$<$14&$<$12&$-2.6_{-0.4}^{+0.4}$\\
19&17:47:52.472&0.16&-24:48:58.47&0.14&3.59&&38$\pm$7&33$\pm$5&26$\pm$7&$<$38&$<$69& &$-0.7_{-0.5}^{+0.4}$\\
20&17:47:50.435&0.06&-24:46:34.66&0.12&3.28&&39$\pm$6&33$\pm$5&22$\pm$6&$<$28&$<$51&$<$122&$-0.9_{-0.5}^{+0.4}$\\
21&17:48:17.922&0.16&-24:46:31.89&0.14&2.97&&26$\pm$7&32$\pm$4&18$\pm$6&$<$23&$<$39&$<$78&$-0.7_{-0.4}^{+0.4}$\\
22&17:48:05.089&0.05&-24:46:45.11&0.10&0.05&Ter5Y  &46$\pm$7&31$\pm$4&$<$13&$<$11&$<$14&$<$12&$-2.2_{-0.5}^{+0.5}$\\
23&17:48:08.549&0.16&-24:44:15.51&0.14&2.62&&$<$19&30$\pm$4&$<$15&$<$18&$<$27&$<$40&$-0.8_{-0.6}^{+0.5}$\\
24&17:48:14.260&0.16&-24:46:51.09&0.14&2.14&&38$\pm$6&30$\pm$4&20$\pm$5&33$\pm$5&$<$22&$<$29&$-0.2_{-0.3}^{+0.3}$\\
25&17:48:15.084&0.16&-24:47:47.33&0.14&2.55&&35$\pm$6&30$\pm$4&23$\pm$5&$<$19&$<$29&$<$48&$-0.9_{-0.4}^{+0.4}$\\
26&17:48:04.870&0.05&-24:46:46.48&0.10&0.03&Ter5I  &$<$20&27$\pm$4&$<$13&$<$11&$<$14&$<$12&$-1.2_{-0.6}^{+0.5}$\\
27&17:48:04.914&0.05&-24:46:53.75&0.10&0.15&Ter5N  &62$\pm$7&26$\pm$4&$<$13&$<$11&$<$14&$<$12&$-3.0_{-0.3}^{+0.5}$\\
28&17:48:03.411&0.16&-24:46:35.59&0.14&0.36&Ter5E  &46$\pm$7&25$\pm$4&$<$13&$<$11&$<$14&$<$12&$-2.5_{-0.6}^{+0.6}$\\
29&17:48:05.087&0.16&-24:46:38.24&0.14&0.12&Ter5F  &41$\pm$7&24$\pm$4&21$\pm$4&$<$11&$<$14&$<$12&$-1.5_{-0.5}^{+0.4}$\\
30&17:48:04.682&0.05&-24:46:51.27&0.10&0.12&Ter5O  &53$\pm$7&23$\pm$4&23$\pm$4&14$\pm$4&$<$14&$<$12&$-1.2_{-0.3}^{+0.3}$\\
31\tablenotemark{b}&17:48:05.015\tablenotemark{c}&0.16&-24:46:43.87\tablenotemark{c}&0.14&0.04&&$<$20&22$\pm$4&32$\pm$3&14$\pm$4&$<$14&$<$12&$-0.2_{-0.2}^{+0.2}$\\
32&17:48:04.726&0.16&-24:46:35.88&0.14&0.15&Ter5L  &30$\pm$7&20$\pm$4&$<$13&$<$11&$<$14&$<$12&$-2.1_{-0.8}^{+0.7}$\\
33&17:48:03.938&0.16&-24:46:47.75&0.14&0.21&Ter5K  &24$\pm$7&18$\pm$4&$<$13&$<$11&$<$14&$<$12&$-1.9_{-0.9}^{+0.8}$\\
34\tablenotemark{d}&17:48:04.677&0.16&-24:46:48.74&0.14&0.08&&$<$20&16$\pm$4&$<$13&$<$11&$<$14&$<$12&$-1.3_{-1.1}^{+0.9}$\\
35&17:48:04.606&0.21&-24:46:42.53&0.23&0.07&CX1&50$\pm$7&$<$12&$<$13&$<$11&$<$14&$<$12&$\leq-1.51$\\
36&17:48:04.789&0.21&-24:46:42.88&0.23&0.03&Ter5ab &53$\pm$7&$<$12&$<$13&$<$11&$<$14&$<$12&$\leq-1.69$\\
37&17:48:10.324&0.16&-24:47:59.24&0.14&1.76&&43$\pm$6&28$\pm$4&31$\pm$5&20$\pm$5&$<$20&$<$23&$-0.6_{-0.3}^{+0.2}$\\
38\tablenotemark{d}&17:48:04.633&0.16&-24:46:46.01&0.14&0.05&&38$\pm$7&16$\pm$4&$<$13&$<$11&$<$14&$<$12&$-2.7_{-0.5}^{+0.7}$\\
39&17:48:05.913&0.21&-24:46:05.83&0.23&0.69&Ter5D  &37$\pm$7&$<$12&$<$13&$<$11&$<$14&$<$13&$\leq-0.66$\\
40\tablenotemark{d}&17:48:04.418&0.16&-24:46:48.78&0.14&0.12&&36$\pm$7&15$\pm$4&$<$13&$<$11&$<$14&$<$12&$-2.7_{-0.5}^{+0.8}$\\
41&17:48:03.760&0.21&-24:47:23.61&0.23&0.70&&31$\pm$7&$<$12&$<$13&$<$11&$<$15&$<$14&$\leq0.08$\\
42&17:48:05.225&0.07&-24:46:47.66&0.12&0.10&EXO 1745&$<$20&$<$12&$<$13&$<$11&99$\pm$4&95$\pm$4&\\
43\tablenotemark{e}&17:48:04.838&0.05&-24:46:42.59&0.10&0.03&Ter5W  &$<$20&$<$12&$<$13&$<$11&$<$14&$<$12&\\
\enddata

\tablenotetext{a}{Other ID for previously known sources. Pulsars Ter5A \citep{1990Natur.347..650L,Nice00} and Ter5P \citep{2005Sci...307..892R, 2006Sci...311.1901H} are redbacks. Pulsar Ter5O is a black widow system \citep{2005Sci...307..892R, 2006Sci...311.1901H}. Ter5 CX1 is a transitional MSP candidate \citep{2018ApJ...864...28B}. EXO 1745--248 is an outbursting neutron star LMXB \citep{2016MNRAS.460..345T}.}
\tablenotetext{b}{Candidate stellar-mass black hole.}
\tablenotetext{c}{This is the weighted mean position across all frequencies (see Sec \ref{sec:bh_can}). A more accurate position may be
the average of the 3.2/3.4/7.4 GHz positions: (R.A., Dec.) = (17:48:05.040, --24:46:43.45). The 5.0 GHz position is: (17:48:04.970, --24:46:44.05).}
\tablenotetext{d}{Candidate redback MSP.}
\tablenotetext{e}{This source is well-detected in the 2014 S band data at exactly the position of the pulsar Ter5 W. Hence we list it in the catalog, even though it is not detected in the 2012 S band data, probably due to confusion and its faintness.}
\tablecomments{The astrometric, flux density, and spectral index uncertainties listed are $1\sigma$ (68\%) quantities, except for the upper limits, which are given at the $3\sigma$ level.}
\end{deluxetable}

\begin{deluxetable*}{llcccc}
\tabletypesize{\footnotesize}
\centerwidetable
\tablewidth{0pt}
\tablecaption{X-ray counterparts to radio sources from Table \ref{tab:radio}.\label{tab:xray}}
\tablehead{
\colhead{\#}  & \colhead{X-ray ID} & \colhead{Alt ID} & \colhead{$\Gamma$} & \colhead{$L_{0.5-10\,\kev}$} & \colhead{Var significance} \\
& & &\colhead{($10^{31}\,\ergs$)} &  (\%)
}
\startdata
1& CXOU J174802.26-244637.5 &Ter5A &$1.5_{-2.0}^{+2.1}$&$0.05_{-0.05}^{+1.29}$&$>$99.9\\
5& CXOU J174805.05-244641.0 &Ter5P &$1.0_{-0.1}^{+0.1}$&$43.8_{-1.6}^{+1.7}$&$>$99.9\\
7& CXOU J174804.53-244634.7 &Ter5C  &$1.5_{-2.0}^{+2.0}$&$0.02_{-0.01}^{+0.12}$&\\
9& CXOU J174805.10-244634.2 &Ter5V  &$3.6_{-0.8}^{+0.4}$&$2.6_{-1.6}^{+3.1}$&\\
17& CXOU J174804.58-244640.5 &Ter5M  &$4.0_{-0.1}^{+0.0}$&$19.7_{-2.7}^{+2.9}$&$>$95.0\\
18& CXOU J174804.96-244645.7 &Ter5Z  &$3.1_{-1.2}^{+0.7}$&$1.9_{-1.0}^{+2.7}$&\\
22& CXOU J174805.09-244644.5 &Ter5Y  &$1.4_{-0.5}^{+0.6}$&$2.28_{-0.40}^{+0.53}$&$>$95.0\\
26& CXOU J174804.87-244646.1 &Ter5I  &$3.1_{-0.5}^{+0.5}$&$9.8_{-3.6}^{+7.4}$&$>$99.9\\
27& CXOU J174804.91-244653.7 &Ter5N  &$2.7_{-2.3}^{+1.0}$&$0.53_{-0.50}^{+2.13}$&\\
28& CXOU J174803.40-244635.4 &Ter5E  &$1.5_{-2.0}^{+2.1}$&$0.02_{-0.01}^{+0.13}$&\\
29& CXOU J174805.11-244638.0 &Ter5F  &$1.6_{-2.0}^{+1.9}$&$0.02_{-0.02}^{+0.17}$&\\
30& CXOU J174804.69-244651.1 &Ter5O  &$1.2_{-0.5}^{+0.5}$&$2.3_{-0.4}^{+0.5}$&\\
31& CXOU J174805.05-244643.1 & Ter5-VLA31  &[$2.1_{-0.5}^{+0.5}$]&[$2.1_{-0.5}^{+0.5}$]&\\
32& CXOU J174804.75-244635.4 &Ter5L  &$1.5_{-2.0}^{+2.0}$&$0.03_{-0.02}^{+0.19}$&\\
33& CXOU J174803.90-244647.7 &Ter5K  &$1.3_{-1.9}^{+2.2}$&$0.02_{-0.02}^{+0.19}$&\\
34& CXOU J174804.69-244648.4 & Ter5-VLA34 &[$1.5_{-0.2}^{+0.2}$]&[$8.0_{-0.8}^{+0.8}$]&$>$99.9\\
35& CXOU J174804.58-244641.8 &CX1&$1.2_{-0.1}^{+0.1}$&$43.9_{-1.6}^{+1.9}$&$>$99.9\\
36& CXOU J174804.75-244642.5 &Ter5ab &$3.9_{-0.1}^{+0.1}$&$10.2_{-1.8}^{+1.7}$&$>$99.9\\
38& CXOU J174804.63-244645.2 & Ter5-VLA38   &[$1.3_{-0.1}^{+0.1}$]&[$25.5_{-1.3}^{+1.3}$]&$>$99.9\\
39& CXOU J174805.92-244605.6 &Ter5D  &$1.1_{-1.8}^{+2.3}$&$0.01_{-0.01}^{+0.07}$&\\
40& CXOU J174804.47-244648.7 & Ter5-VLA40 &[$1.3_{-0.7}^{+1.0}$]&[$1.5_{-0.4}^{+0.9}$]&$>$99.0\\
42& CXOU J174805.23-244647.3 &EXO\,1745&$1.5_{-0.0}^{+0.0}$&$179_{-3}^{+3}$&$>$99.9\\
\enddata
\tablecomments{Values in square brackets indicate
the refit parameters for the new sources (see Sec \ref{sec:data_analysis}). All remaining values are from \citet{goose}.} Uncertainties on X-ray luminosity and photon index are at the 90\% level.
\end{deluxetable*}

\end{document}